\documentclass[twocolumn,english,aps,pra,10pt,superscriptaddress,floatfix]{revtex4-2}
\usepackage{times}
\usepackage{graphicx}
\usepackage{amssymb}
\usepackage{bm}
\usepackage{amssymb,amsfonts,amsmath,amsbsy,bm,t1enc,latexsym}
\usepackage{float}
\usepackage[colorlinks=true,citecolor=blue,linkcolor=magenta]{hyperref}
\usepackage[markup=blue, authormarkupposition=left]{changes}
\usepackage[english]{babel}
\usepackage{url}
\usepackage{siunitx}
\DeclareSIUnit \dbc {dBc}
\usepackage{soul}
\usepackage{changes}
\usepackage{cancel}
\usepackage{times}

\newcommand{\SiN }[0]{Si$_3$N$_4$}

\begin{document}
	
	\title{Narrow-linewidth, piezoelectrically tunable photonic integrated blue laser}

	\author{Anat Siddharth}
	\affiliation{Laboratory of Photonics and Quantum Measurements, Swiss Federal Institute of Technology Lausanne (EPFL), CH-1015 Lausanne, Switzerland}
	\affiliation{Center of Quantum Science and Engineering, EPFL, CH-1015 Lausanne, Switzerland}
	
	\author{Asger B. Gardner}
	\affiliation{Aarhus University, 8200 Aarhus N, Denmark}	
	
	\author{Xinru Ji}
	\affiliation{Laboratory of Photonics and Quantum Measurements, Swiss Federal Institute of Technology Lausanne (EPFL), CH-1015 Lausanne, Switzerland}
	\affiliation{Center of Quantum Science and Engineering, EPFL, CH-1015 Lausanne, Switzerland}
	
	\author{Shivaprasad U. Hulyal}
	\affiliation{Laboratory of Photonics and Quantum Measurements, Swiss Federal Institute of Technology Lausanne (EPFL), CH-1015 Lausanne, Switzerland}
	\affiliation{Center of Quantum Science and Engineering, EPFL, CH-1015 Lausanne, Switzerland}
		
	\author{Mikael S. Reichler}
	\affiliation{Laboratory of Photonics and Quantum Measurements, Swiss Federal Institute of Technology Lausanne (EPFL), CH-1015 Lausanne, Switzerland}
	\affiliation{Center of Quantum Science and Engineering, EPFL, CH-1015 Lausanne, Switzerland}
    
    \author{Alaina Attanasio}
	\affiliation{Purdue University, West Lafayette, IN-47901, United States}
	
	\author{Johann Riemensberger}
	\affiliation{Laboratory of Photonics and Quantum Measurements, Swiss Federal Institute of Technology Lausanne (EPFL), CH-1015 Lausanne, Switzerland}
	\affiliation{Center of Quantum Science and Engineering, EPFL, CH-1015 Lausanne, Switzerland}
	
	\author{Sunil A. Bhave}
	\affiliation{Purdue University, West Lafayette, IN-47901, United States}
	
	\author{Nicolas Volet}
	\affiliation{Aarhus University, 8200 Aarhus N, Denmark}	
	
	\author{Simone Bianconi}
	\email[]{simone.bianconi@epfl.ch}
	\affiliation{Laboratory of Photonics and Quantum Measurements, Swiss Federal Institute of Technology Lausanne (EPFL), CH-1015 Lausanne, Switzerland}
	\affiliation{Center of Quantum Science and Engineering, EPFL, CH-1015 Lausanne, Switzerland}
	
	\author{Tobias J. Kippenberg}
	\email[]{tobias.kippenberg@epfl.ch}
	\affiliation{Laboratory of Photonics and Quantum Measurements, Swiss Federal Institute of Technology Lausanne (EPFL), CH-1015 Lausanne, Switzerland}
	\affiliation{Center of Quantum Science and Engineering, EPFL, CH-1015 Lausanne, Switzerland}
	
	\medskip
	\maketitle
	
\noindent\textbf{Frequency-agile lasers operating in the ultraviolet-to-blue spectral range (360–480 nm) are critical enablers for a wide range of technologies, including free-space and underwater optical communications, optical atomic clocks, and Rydberg-atom-based quantum computing platforms. Integrated photonic lasers offer a compelling platform for these applications by combining low-noise performance with fast frequency tuning in a compact, robust form factor through monolithic integration. However, realizing such lasers in the blue spectral range remains challenging due to limitations in current semiconductor materials and photonic integration techniques.
Here, we report the first demonstration of a photonic integrated blue laser at around 461 nm, which simultaneously achieves frequency agility and low phase noise. This implementation is based on the hybrid-integration of a gallium nitride (GaN)-based laser diode, which is self-injection locked to a high-Q microresonator fabricated on a low-loss silicon nitride (\SiN) photonic platform with $\sim 0.4$ dB/cm propagation loss. The laser exhibits a sub-30 kHz linewidth and delivers over 1 mW of optical output power. In addition, aluminum nitride (AlN) piezoelectric actuators are monolithically integrated onto the photonic circuitry to enable high-speed modulation of the refractive index, and thus tuning the laser frequency. This enables mode-hop-free laser linear frequency chirps with excursions up to 900 MHz at repetition rates up to 1 MHz, with tuning non-linearity below 2\%. 
We showcase the potential applications of this integrated laser in underwater communication and coherent aerosol sensing experiments.
}
	
		\begin{figure*}[htbp!]
		\centering
		\includegraphics[width=1\linewidth]{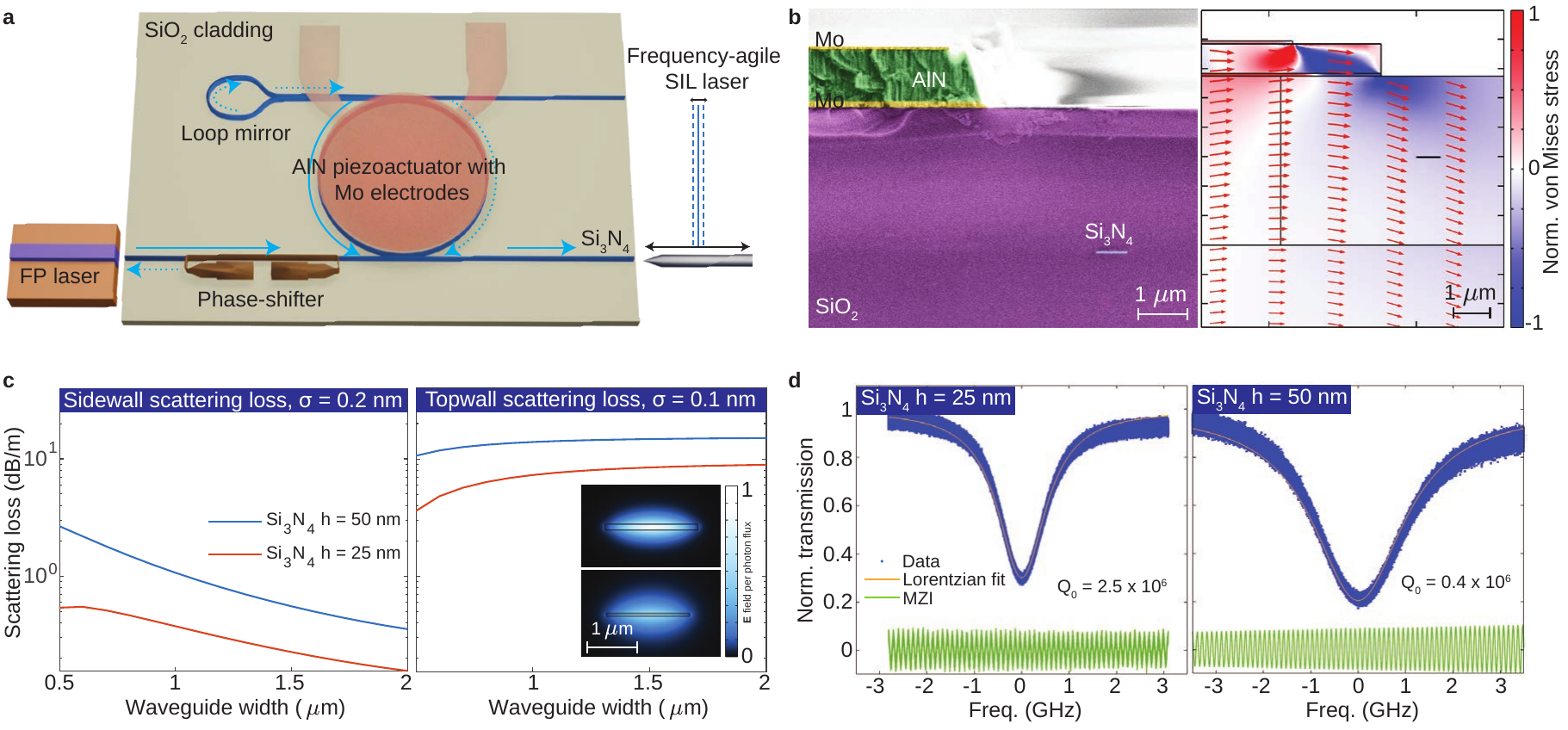}
		\caption{ \textbf{Low-noise, FMCW photonic integrated blue laser.}
(a) Schematic of the chip-scale laser architecture, showing a frequency-agile self-injection locked (SIL) laser loop based on a high-Q \SiN microring resonator integrated with an AlN piezoelectric actuator for high-speed tuning.
(b) Cross-sectional SEM of the fabricated device (left) showing AlN and \SiN layers, and FEM simulation of stress-induced refractive index modulation from the AlN piezoactuator (right), illustrating the strain profile influencing tuning efficiency.
(c) Simulated waveguide scattering loss as a function of waveguide width for \SiN core thicknesses of 25 nm and 50 nm. Sidewall ($\sigma$ = 0.2 nm) and topwall ($\sigma$ = 0.1 nm) roughness are analyzed separately, $\sigma^2$ is the mean square roughness. Insets show transverse mode profiles for different geometries.
(d) Measured transmission spectra and Lorentzian fits for microring resonators fabricated with 25~nm and 50~nm \SiN cores. The extracted loaded quality factors are 2.5$\times$10$^6$ and 0.4$\times$10$^6$, respectively, confirming the advantage of thinner cores in achieving ultra-low-loss performance.}
\label{Fig:fig1}
		
	\end{figure*}

	
\section*{Introduction}

Recent advancements in quantum technologies and photonics have fueled a growing demand for laser sources operating in the ultraviolet (UV) and blue (360–480 nm) spectral range. These wavelengths are essential for a wide spectrum of applications, including high-precision timekeeping, Doppler laser cooling of atoms and ions, quantum state manipulation, and entangled photon pair generation for quantum networks \cite{wineland1983laser,Chu1997,Weihs2014,Anwar2020,Weiss2025}. 
Beyond quantum systems, UV-blue lasers have attracted increasing attention for atmospheric sensing of aerosols \cite{mcmurry2000review} and underwater optical ranging and communication \cite{filisetti2018developments,mcleod2013autonomous} due to their stronger scattering in air ($\propto \lambda^{-4}$) and lower attenuation in water at shorter wavelengths.
Among the most accurate frequency standards developed to date are optical atomic clocks, which rely on ultra-narrow optical transitions in alkaline-earth elements such as strontium (Sr) and ytterbium (Yb).
Both neutral atom and ion-based clock configurations have demonstrated remarkable performance in precision measurement and fundamental physics tests \cite{poli2014transportable,zhang2022ytterbium}.
The crucial electronic transitions required for laser cooling and state preparation in these atoms occur within the UV-VIS spectrum, specifically at 369 nm (Yb$^+$), 399 nm (Yb), 421 nm (Sr$^+$), and 461 nm (Sr) \cite{nicholson2015systematic,ishiyama2023observation}.
With regards to applications related to quantum information processing, Rydberg-atom-based quantum computing platforms rely on narrow-linewidth, low-noise blue lasers for high-fidelity excitation of highly excited atomic states, enabling fast and coherent two-qubit gate operations \cite{radnaev2024universal,levine2018high}.

Currently, most laser systems serving these roles are based on semiconductor diode lasers, particularly Fabry-Pérot (FP) designs. However, FP laser diodes typically operate in multimode regimes and require external stabilization to achieve the single-frequency, narrow-linewidth performance necessary for metrology and quantum applications. This is commonly realized through external free-space cavities and gratings \cite{ricci1995compact}, which introduce complexity and alignment sensitivity. These constraints and the size of such systems limit their portability and environmental robustness.
To overcome these limitations, recent efforts have explored the implementation of photonic integrated lasers at blue wavelengths \cite{siddharth2022near,Corato-Zanarella2022,franken2024widely}. 
However, present implementations lack fast wavelength tuning mechanisms for frequency chirping, a capability that is critical for active frequency stabilization (i.e locking via PDH technique \cite{black2001introduction}) and dynamic applications such as coherent sensing and underwater optical communications.

In this work, we address the limitations of existing photonic platforms and demonstrate an integrated laser uniquely combining low-noise performance (narrow linewidth $\mathcal{O}$(kHz)) with fast frequency tuning (fast chirping $\mathcal{O}$(MHz)).
We introduce a low-loss photonic integrated circuit (PIC) architecture based on weakly confining 25~nm thick \SiN~waveguides. Compared to similar waveguides manufactured in a 50 nm-thick \SiN, the thinner 25 nm-thick \SiN~waveguides feature a reduced optical confinement that enables significantly reducing scattering losses, demonstrating resonators with intrinsic quality factors exceeding $2.5 \times 10^6$, corresponding to $0.4$ dB/cm propagation loss.
In addition, these platforms incorporate monolithically integrated aluminum nitride (AlN) piezoelectric actuators to enable high speed and energy efficient modulation of the refractive index. 
Leveraging this architecture, we demonstrate a hybrid integrated low-noise laser system in which a gallium nitride (GaN)-based laser diode is self-injection locked to high-Q \SiN microresonators. 
The resulting laser exhibits kilohertz-level linewidth and fast tunability, positioning it as a highly compelling solution for emerging applications in quantum information processing, precision sensing, and underwater optical communications.

\section*{Low-loss photonic platform for visible integrated photonics}

The hybrid photonic integrated laser architecture comprises a Fabry-Pérot (FP) laser diode which is butt-coupled to a \SiN photonic integrated cavity, as illustrated in Figure~\ref{Fig:fig1}a. The \SiN~PIC includes a high-Q microring resonator featuring a drop waveguide with a Sagnac mirror for facilitating self-injection locking (SIL) by accurately controlling the back-reflection into the FP laser diode \cite{galiev2021mirror}. 
The microresonator frequencies can be modulated via the stress-optic effect using a monolithically integrated AlN piezoelectric actuator for voltage-controlled tuning of the laser frequency.
The FP laser diode is butt-coupled to the chip via a horn-tapered waveguide designed to optimize mode overlap and minimize insertion loss. The laser chip is mounted on a thermoelectric cooler (TEC) to maintain stable operation at 25$^\circ$C.

The device cross-section is shown in the false-colored SEM image in Figure~\ref{Fig:fig1}b, revealing the multilayer stack comprising the \SiN~waveguide core embedded in SiO\textsubscript{2} cladding. The bottom oxide is a 4 $\mu$m thermal oxide layer, while the top oxide consists of TEOS and low-temperature oxide (LTO) with a combined thickness of 3 $\mu$m. The entire photonic stack sits on a 230 $\mu$m-thick silicon substrate. On top of the upper cladding, the AlN piezoelectric film and molybdenum electrodes are patterned, enabling localized stress application to the resonator \cite{siddharth2024piezoelectrically}. Finite element simulations, shown on the right, confirm that the induced mechanical stress results in a tunable refractive index shift via the stress-optic effect, facilitating high-speed tuning of the cavity resonance \cite{liu2020monolithic}.

To investigate the impact of waveguide geometry on optical loss, we fabricated two versions of the PIC platform, with core \SiN~waveguide thicknesses of 25 nm and 50 nm, both with a fixed width of 800 nm. The rationale for pursuing ultra-thin waveguide geometries is to leverage low mode confinement to reduce scattering losses \cite{bauters2011ultra}.
Using the Lacey-Payne model \cite{payne1994theoretical}, we simulate scattering losses from sidewall roughness ($\sigma$ = 0.2 nm) and topwall roughness ($\sigma$= 0.1 nm) for both thicknesses over a range of waveguide widths. In both cases, the 25 nm platform exhibits significantly lower losses, as shown in Figure~\ref{Fig:fig1}c. This is attributed to the low-confinement modes in the thinner waveguides, which reduce the electric field intensity at scattering-inducing boundaries, as illustrated in the inset of Figure~\ref{Fig:fig1}c (cf. Supplementary Notes).

Experimental evidence confirms this trend: the thinner 25 nm platform enables microring resonators with a significantly higher optical quality factor (Q), as showcased in Figure~\ref{Fig:fig1}d. Lorentzian fits to the transmission spectra yield intrinsic Q-factors of $2.5 \times 10^6$ for the 25 nm waveguide and $4 \times 10^5$ for the 50 nm counterpart, demonstrating more than a sixfold improvement in cavity coherence thanks to the reduction in scattering losses.
This substantial improvement is especially critical for the noise characteristics of the self-injection-locked laser, given that the linewidth reduction factor scales with the Q-factor \cite{ilchenko2011compact}.
A comparison of intrinsic Q across different material platform in the UV-blue spectral range is provided in the Supplementary notes.

\section*{Self-injection locked blue laser operation and fast frequency tuning}

\begin{figure*}[htb!]
		\centering
		\includegraphics[width=1\linewidth]{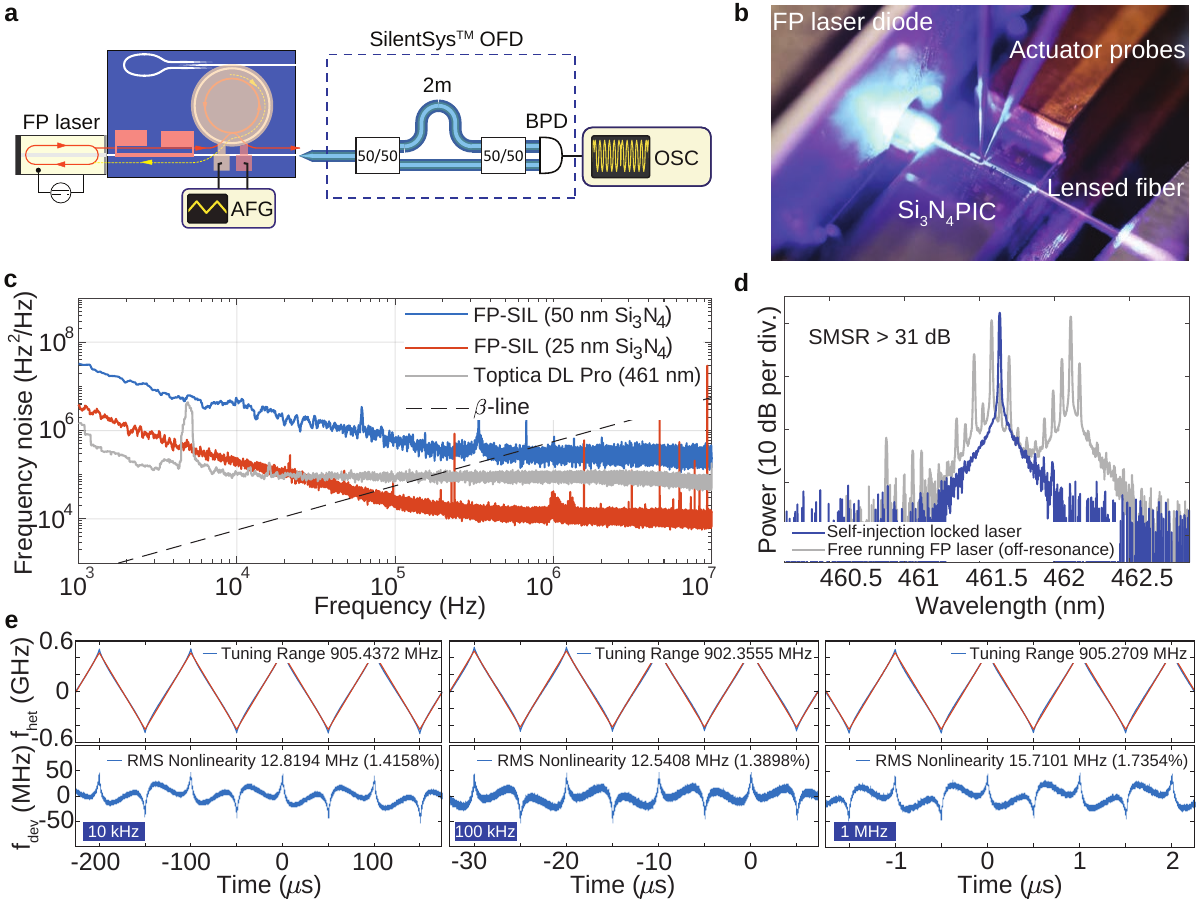}
		\caption{ \textbf{Self-injection locked blue laser characterization.}
            (a) Schematic of the experimental setup to measure the frequency noise and linearity analysis of the tuning of self-injection locked laser. FP: Fabry-Perot, AFG: Arbitrary Function Generator, OFD: Optical Frequency Discriminators, BPD: balanced photodetector, OSC: oscilloscope.
            (b) Photograph of the experimental setup showing the hybrid integration of the FP laser diode and the photonic-MEMS chip to realize self-injection locking.
            (c) Frequency noise spectra of the photonic integrated laser at the wavelength of 461.5 nm. FP-SIL: Fabry-Perot self-injection locked.
            (d) Optical spectra of the photonic integrated laser at the wavelength of 461.5 nm.
            (e) Time-frequency dependence of the FMCW signal obtained from homodyne beatnotes for different chirp frequencies. The bottom row shows the residual of the least squares fitting of the experimental time-frequency traces with a linear triangular chirp pattern.}
        \label{Fig:fig2}
\end{figure*}

The realization of the low-loss \SiN~platform, with propagation losses below 1 dB/cm at 461 nm discussed above, is essential for enabling a self-injection-locked photonic integrated laser with low noise performance. Indeed, the linewidth reduction granted by self-injection locking of a laser diode to a high-Q cavity in a weak feedback regime is proportional to the square of the two quality factors \cite{kondratiev2017self}:

\begin{equation}
\frac{\delta \omega}{\delta \omega_{\text{free}}} \approx \frac{Q_{\text{DFB}}^2}{Q^2} \cdot \frac{1}{16R(1 + \alpha_g^2)},
\end{equation}
where $\delta \omega_{\text{free}} / 2\pi$ is the linewidth of the free-running FP laser; $\delta \omega / 2\pi$ is the linewidth of the self-injection-locked DFB laser; $Q_{\text{DFB}}$ and $Q = \omega / \kappa$ are the quality factors of the laser diode cavity and of the microresonator mode, respectively (with $\kappa = \kappa_{\text{ex}} + \kappa_0$, where $\kappa_0$ and $\kappa_{\text{ex}}$ are the intrinsic cavity decay rate and bus–waveguide coupling rate, respectively); and $\alpha_g$ is the phase–amplitude coupling factor of the laser diode.


Figure~\ref{Fig:fig2}b shows a photograph of the self-injection locked laser in operation, highlighting the hybrid integration of the FP laser diode with the \SiN-AlN MEMS-photonic chip, including a lensed fiber for output coupling and actuator probes.
To evaluate the spectral purity and frequency stability of the SIL laser, we employed an Optical Frequency Discriminator (OFD) system by SilentSys$^\text{TM}$, combined with acquisition on an oscilloscope (OSC), as shown in Figure~\ref{Fig:fig2}a. 
The OFD system consists of a 2-meter delay line interferometer and a balanced photodetector (BPD). This system allows for precise characterization of both the frequency noise spectrum and the tuning dynamics.
Figure~\ref{Fig:fig2}c presents the single sideband power spectral density (PSD) of the frequency noise for the SIL lasers fabricated using 25 nm and 50 nm thick \SiN.
For comparison, we also include the frequency noise PSD of a commercial Toptica DL Pro 461 nm external cavity diode laser.
The SIL laser based on the 25 nm-thick \SiN~platform demonstrates a marked improvement in frequency noise performance compared to both the 50 nm device and the commercial laser, reaching an intrinsic noise floor of approximately $10^5$ Hz\textsuperscript{2}/Hz at a around 4 MHz offset, corresponding to an intrinsic Lorentzian linewidth of about 330 kHz. 
This highlights the benefit of reduced modal overlap with scattering boundaries in the ultrathin waveguide platform, which leads to lower phase noise and higher coherence.
The optical spectrum shown in Figure~\ref{Fig:fig2}d reveals a strong central mode at 461.5 nm with a side mode suppression ratio (SMSR) greater than 31 dB when locked to the \SiN~resonator. The spectrum of the free-running FP laser is also shown for comparison, to showcase the significant suppression of multimode behavior enabled by self-injection locking.
The thinner \SiN~core improves coupling efficiency both from the FP laser to the microresonator and from the microresonator to the lensed fiber.
This enhanced mode matching results in significantly higher output power in the self-injection locked configuration.
Specifically, with the 25 nm thick \SiN~waveguide, the output power reaches approximately 2~mW, compared to only 0.8~mW for the 50 nm thick \SiN, under identical drive conditions of 80 mA applied to the FP laser.

The monolithically integrated AlN piezoelectric actuators enable to apply a voltage to control the refractive index of the \SiN~waveguides in the microresonator via the stress optic effect as shown in Figure~\ref{Fig:fig1}b.
We demonstrated fast laser frequency tuning with linear frequency chirps by applying a symmetric sawtooth voltage waveform with 50 V peak-to-peak and various repetition rates up to 1 MHz to the piezoelectric actuators' electrodes as shown in Figure~\ref{Fig:fig2}e. 
We investigated piezoelectric frequency tuning of integrated lasers based on both the 25 nm and 50 nm-thick \SiN platforms, the latter are shown in Figure~\ref{Fig:fig2}e. 
The upper panels show the evolution of frequency versus time, extracted from the output of a Mach Zehnder interferometer using a short time Fourier transform (STFT). At all repetition rates, the tuning range exceeds 900 MHz with a nonlinearity below 2 \%.
The lower panels quantify the deviation from ideal triangular frequency modulation, yielding root mean square (RMS) nonlinearities between 1.4~\% and 1.7~\%, all achieved without any pre-distortion or active compensation.

The frequency excursions of the chirps for the 50 nm-thick \SiN~platform shown in Figure~\ref{Fig:fig2}e is higher (900 MHz) than for the 25 nm-thick one (125 MHz), shown in the Supplementary notes.
The tuning efficiency for the 50 nm thick \SiN~SIL laser is 18 MHz/V and for the 25 nm thick \SiN~SIL laser is 12.4 MHz/V. 
This is due to the choice of maintaining the waveguide geometries identical in the two PIC platforms to ensure a fair comparison between them: the lower confinement of the thinner 25 nm-thick \SiN~waveguides results in a higher bending loss for a given bend radius. This detrimentally affects the reflectivity of the Sagnac mirror, which in turn limits the self-injection locking range, as discussed in the Supplementary notes. 
This reduced locking range causes laser to lose the SIL state during the chirp at a lower excursion compared to what is supported by the 25 nm-thick \SiN~PIC.
As a result, the photonic integrated blue lasers presented here can operate with either a sub-30 kHz linewidth and frequency tuning excursions of 125 MHz, or with sub-600 kHz linewidth and frequency tuning excursions of 900 MHz. Nonetheless, this performance can be readily improved upon by optimizing the bending radius of the 25 nm-thick \SiN~waveguides to enable simultaneous operation with sub-30 kHz linewidth and frequency tuning excursions of 900 MHz.

Finally, the linearity of the piezoelectric frequency tuning can be further improved through hermetic hybrid packaging, which would mitigate acoustic perturbations and improve long term frequency stability \cite{siddharth2025ultrafast,voloshin2024monolithic}.

\section*{Underwater coherent communication}

\begin{figure*}[ht!]
		\centering
		\includegraphics[width=1\linewidth]{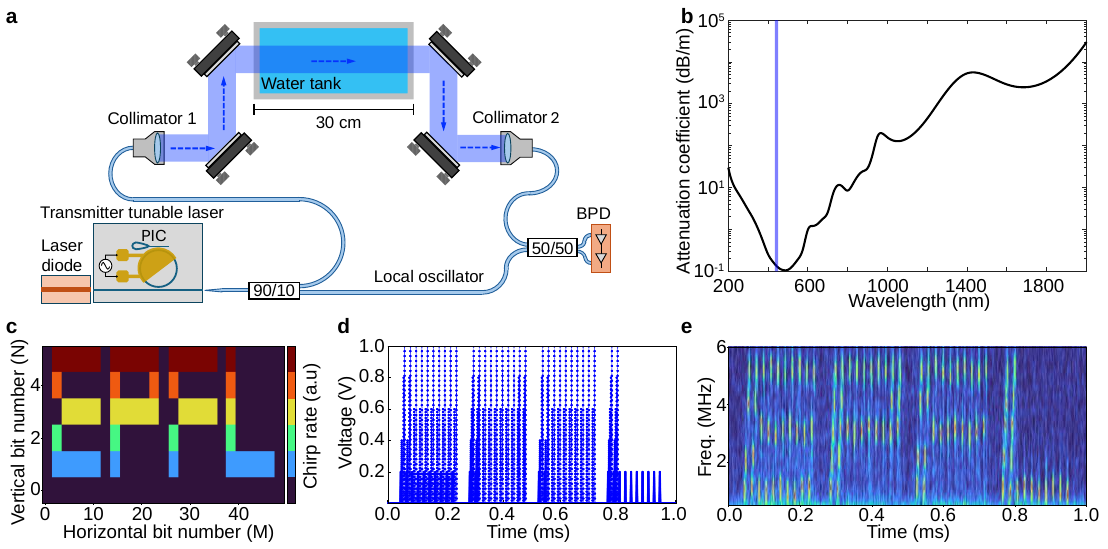}
		\caption{ \textbf{Underwater FMCW communication link using the integrated blue laser.} (a) Schematic of the underwater communication setup. The frequency-chirped  output of the laser is split into two paths, one which passes through 30 cm of water while the other is kept in fiber. The two paths are joined in a 50:50 splitter and passed to a balanced photodetector (BPD) to detect the beatnote. (b) Absorption spectrum of water. The blue line signifies the area around 450 nm. (c) Encoding of the EPFL logo onto a 6-level system in frequency space by discretized laser chirp rates. (d) Encoding of the logo from (c) into an AFG voltage-time waveform imparted onto the laser piezoactuators. (e) Reconstruction of the logo through the short-time Fourier transform of the signal obtained from the BPD.}
        \label{Fig:fig3}
\end{figure*}

The unique combination of low phase noise and rapid frequency agility demonstrated by the photonic integrated blue laser makes it exceptionally well suited for coherent communications and sensing applications, where precise frequency control and spectral purity are critical. These coherent techniques are especially attractive because they can leverage intrinsic background noise rejection and robust encoding methods. Finally, the compact form factor is suitable for field-deployable applications \cite{filisetti2018developments,mcleod2013autonomous}.

An especially compelling application of this technology is point-to-point underwater communication: this is because the absorption of water at blue wavelengths is several orders of magnitude below that at telecom wavelengths (see Figure~\ref{Fig:fig3}b), thereby enabling light propagation over tens to hundreds of meters with an extinction of only a few dB (0.1 dB/m).
We demonstrated a coherent underwater communication experiment using the setup illustrated in Figure~\ref{Fig:fig3}a. In this homodyne communications demonstration, the frequency-modulated output of the SIL laser is divided into two optical paths using a 90/10 splitter: one path is transmitted in free space through a 30 cm column of water, while the other serves as a reference local oscillator (LO). In a realistic application, this LO would be provided by a different laser deployed onboard the receiver platform and locked to a stable reference, as common in free-space coherent communication. After transmission, both paths are recombined using a 50/50 splitter and directed into a balanced photodetector (BPD), which detects the resulting heterodyne beatnote signal. Due to the difference in propagation length between the two arms, a frequency chirp is transduced into a heterodyne beatnote which depends on the chirp rate and the length of the delay, in a manner similar to frequency-modulated continuous wave (FMCW) techniques. This enables to encode a 6-bit sequence into different frequency chirps, which can be detected and demodulated using a local oscillator, as detailed in the Methods section. Here, the fast frequency chirp capabilities of the photonic integrated SIL laser are key to high-data rates communication and robust information encoding via frequency chirp modulation. As shown in Figure~\ref{Fig:fig3}c, we implement a six-level frequency-shift keying (FSK) protocol, in which the chirp rate of the laser is varied discretely from 0 to 275 THz/s to encode digital information in time bins of 3.3 \textmu s. For the purpose of demonstration, the EPFL logo is encoded using distinct chirp rates, as shown in Figure~\ref{Fig:fig3}d, which displays the corresponding voltage waveform applied to the piezoelectric actuators. 
In Figure~\ref{Fig:fig3}e, the recorded time–frequency response from the BPD is shown via short-time Fourier transform (STFT), clearly reconstructing the encoded chirp pattern and validating accurate and dense data transmission.
This experiment underscores the feasibility of using the frequency-agile blue laser for underwater coherent communication links with MHz-bandwidth frequency modulation and low spectral distortion.

\section*{Aerosol sensing}

\begin{figure*}[ht!]
		\centering
		\includegraphics[width=1\linewidth]{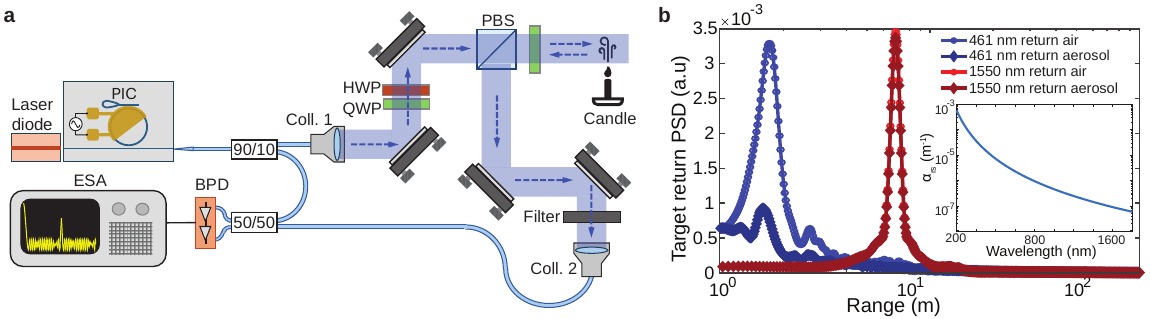}
		\caption{ \textbf{Aerosol LiDAR sensing using the integrated blue laser.} (a) Schematic of the experimental setup for aerosol LiDAR sensing: the frequency-chirped output of the photonic integrated laser is split into two paths, one of which is routed to a free space circulator constructed using a combination of a polarization-dependent beam splitter and wave plates. The return light is coupled back into a single-mode fiber for coherent detection of the beat note with the local oscillator. ESA: electrical spectrum analyzer, BPD: balanced photodetector, PBS: polarization-dependent beam splitter, HWP: half-wave plate, QWP: quarter-wave plate. (b) Aerosol sensing is performed in an FMCW LiDAR scheme, where the transmitted signal is reflected off a distant reflector, where the aerosol scatterer (in this case the smoke from a candle) is distributed along the beam path. The FMCW beat note is attenuated by the aerosol scatterer along that path as shown in the beat note comparison. A similar setup with equivalent architecture was simultaneously used for comparing to the aerosol sensitivity at 1550 nm with the same beam path, showing no detectable change for the case of the 1550 nm optical signal. The inset shows the dependency of Rayleigh scattering coefficient ($\alpha_{\text{rs}}$) on wavelength, showing the utility of shorter wavelength light in aerosol sensing.}
        \label{Fig:fig4}
\end{figure*}

Beyond communications, frequency agile lasers operating at short wavelengths are ideally suited for coherent atmospheric sensing, particularly for detecting small-scale aerosols \cite{baars2017target,veselovskii2005information}. The Rayleigh scattering cross-section scales as $\lambda^{-4}$, making blue and UV wavelengths far more sensitive to aerosol particles than traditional infrared laser-based systems \cite{cimini2010integrated}. 
We demonstrated aerosol sensing of aerosols using the mono-static architecture depicted in Figure~\ref{Fig:fig4}a: this system leverages coherent detection to extract information on the distance traveled by the transmitted beam reflected on a target, which is encoded in the detected heterodyne beatnote like in an FMCW ranging measurement \cite{siddharth2025ultrafast}, allowing to measure attenuation with high dynamic range.
The output of the frequency-modulated SIL laser is emitted into free space and reflected from a target (a white surface placed beyond a light-scattering medium). The return signal is routed via a free-space circulator implemented with a polarization-dependent beam splitter (PBS), half-wave plates (HWP), and quarter-wave plates (QWP), before being coherently combined with a local oscillator, tapped from the same laser output. The heterodyne beatnote detected on a balanced photodetector (BPD) encodes the distance information into frequency and the attenuation information into intensity. This coherent technique crucially allows to distinguish attenuation from aerosol scattering from interfering agents blocking the beam transmission, by simply detecting the heterodyne frequency to validate the distance to the target. We demonstrate the ability of the system to sense aerosol concentration by analyzing the retrieved signal when the free-space path is traversed by an aerosol (candle smoke).
Figure~\ref{Fig:fig4}b compares the LIDAR return power spectral densities (PSDs) under two conditions: with and without candle smoke introduced into the beam path.
Using the blue photonic integrated SIL laser, a substantial reduction in the peak return signal frequency is observed when the aerosol is present, indicating high sensitivity to fine particulate scattering.
We compared these results to measurements performed in the same conditions using a laser operating at 1550 nm, which shows negligible change in the detected signal, confirming its insensitivity to aerosol scatterers.
This comparison highlights the distinct advantage of using blue-wavelength lasers for full-plume aerosol detection, ideal for applications such as environmental studies, combustion process monitoring, and atmospheric particle detection.

\section*{Conclusion}

We have demonstrated a frequency-agile photonic integrated blue laser operating near 461 nm that simultaneously achieves frequency agility, low phase noise, and high spectral purity. This laser was realized through the hybrid integration of a GaN-based laser diode self-injection locked to a high-Q microresonator fabricated on a low-loss \SiN photonic platform, achieving a sub-30 kHz linewidth with over 1 mW of optical output power. These photonic integrated circuits take advantage of low modal confinement in 25 nm-thick \SiN waveguides to significantly reduce scattering losses and drastically enhance the coherence of the self-injection locked laser. We demonstrated the effectiveness of this approach by comparing the losses and laser performance to those obtained with a 50 nm-thick \SiN waveguides. 

The monolithic integration of AlN piezoelectric actuators enables high-speed, mode-hop-free frequency tuning, supporting linear chirps at repetition rates up to 1 MHz, with tuning nonlinearities below 1.5\%. The frequency excursions of the chirps is higher for the 50 nm-thick \SiN platform (900 MHz) than for the 25 nm-thick one, due to the reduced reflection from the Sagnac mirror limiting the self-injection locking range, a feature that can be readily improved upon by optimizing the bending radius of the 25 nm-thick \SiN waveguides.

We showcased the application of this laser with experimental demonstrations of underwater coherent communications and atmospheric sensing.
These results highlight the viability of compact, high-performance blue integrated lasers for emerging applications in coherent communication and sensing, paving the way for broader adoption of integrated photonics in the visible spectrum.

\section*{Methods}

\subsection*{Photonic integrated circuit fabrication}
For the fabrication of passive, ultra-low loss \SiN photonic integrated circuits, a single-crystalline 100 mm Si wafer was wet-oxidized with 8.0~$\mu$m SiO$_2$ to provide a bottom cladding, which can effectively isolate light confined in the thin \SiN waveguide from the silicon substrate, including at edge couplers.
The waveguides are formed from 25~nm thick low-pressure chemical vapor deposition (LPCVD) Si$_3$N$_4$ films.
The LPCVD Si$_3$N$_4$ films exhibit an RMS roughness of 0.3~nm and uniformity of $\pm0.6\%$ across a 4-inch wafer.
Following deposition, Si$_3$N$_4$ films are annealed at 1200$^\circ$C for 11 hours to eliminate excess H$_2$ and break N-H and Si-H bonds.
The waveguide patterns are defined using deep ultraviolet (DUV) stepper lithography (KrF 248~nm) with a resolution of 180~nm.
Si$_3$N$_4$ waveguides are formed via anisotropic dry etching with CHF$_3$ and SF$_6$ gases, yielding vertical, clean, and smooth sidewalls while minimizing polymer re-deposition from etching byproducts.
A 3~$\mu$m-thick hydrogen-free, low-loss SiO$_2$ cladding was subsequently grown at 300$^\circ$C using inductively coupled plasma-enhanced chemical vapor deposition with SiCl$_4$ and O$_2$ as precursors.
The AlN actuators are then fabricated with the same process used in \cite{siddharth2024piezoelectrically}.
After a fully wafer-scale fabrication process, the wafer was separated into individual dies containing chip-scale lasers via deep etching of SiO$_2$ and subsequent deep reactive ion etching (DRIE) of Si using the Bosch method, followed by backside grinding.

\subsection*{Underwater coherent communication and aerosol sensing}
To realize underwater coherent communication using the photonic integrated blue laser, data was encoded using 6-level frequency-shift keying (FSK) as indicated in Figure \ref{Fig:fig3}c, with the full encoding scheme consisting of 6 vertical bits and 50 horizontal bits. Each bit was assigned a fixed chirp rate with faster chirp rates resulting in higher beat notes in the homodyne detection scheme, allowing for multi-level information encoding in frequency space. Through column-wise flattening of the array, the encoding in Figure \ref{Fig:fig3}c was translated into a single 300 bin array, with each bin being assigned a discrete chirp rate. To experimentally realize the encoding with a 1 kHz repetition rate, each bin in the array is assigned a 3.3 µs time slot in the voltage-time waveform. By changing the maximum excursion of the voltage ramp between different discrete levels within each time slot, the chirp rate encoding is translated into a time-dependent voltage signal that can be imparted on the laser. The resulting waveform using this method is given in Figure \ref{Fig:fig3}d. 

To ensure efficient tuning, a PIC on the 50 nm \SiN platform was used for the experiment. A Fabry-Pérot (FP) laser diode with a center wavelength of 450 nm was edge-coupled to the PIC as a gain chip. The 1 kHz waveform in Figure \ref{Fig:fig3}d was uploaded to an arbitrary function generator (AFG) and amplified 50 times by a high-voltage amplifier before being imparted on the piezoactuators using RF probes. The resulting modulated optical signal was collected with a fiber and passed to a 90/10 splitter. The 90 \% signal was coupled into free-space using a collimator, passed through a 30 cm long water column, and collected into fiber by a second collimator. The 10 \% signal was kept in fiber as a local oscillator. The two signals were combined in a 50/50 splitter before being passed to a balanced photodetector (BPD), and the resulting signal was collected on an oscilloscope. To reconstruct the encoded pattern, short-time Fourier transformation (STFT) is employed. The STFT method performs a Fourier transformation of the time-dependent signal in a bin-by-bin fashion, accurately reconstructing the encoded pattern as seen in Figure \ref{Fig:fig3}e. 
  	
\vspace*{10pt}
    
\begin{footnotesize}
		
		
		\noindent \textbf{Author Contributions}:
		A.S. and S.H. designed the photonic integrated circuits. A.S. performed the photonics and stress-optics simulations with help from J.R and M.R. X.J. and A.A. fabricated the samples. A.S., A.B.G and S.B. performed the characterization and experiments and analyzed the data.
		T.J.K., S.A.B., N.V. and S.B. supervised the project.
		
		\noindent \textbf{Acknowledgements}:
		The authors acknowledge Andrea Bancora for his help and expertise in building the free space sensing setup. We thank Antonella Ragnelli for her invaluable support in securing and managing the project funding
				
		\noindent \textbf{Funding Information and Disclaimer}: This material is based on research sponsored by the Army Research Laboratory under agreement number PNT-22-24-024/W911NF-22-2-0050, the EU’s Horizon Europe research and innovation programme under grant No. 101131069 (AgiLight), and the Swiss National Science Foundation under grant agreement No. 211728 (BRIDGE).
		Research at Purdue University was sponsored by the Army Research Laboratory with SEMI-PNT and was accomplished under Cooperative Agreement Number 911NF-22-2-0050.
The views and conclusions contained in this document are those of the authors and should not be interpreted as representing the official policies, either expressed or implied, of the Army Research Laboratory or the U.S. Government or SEMI. The U.S. Government is authorized to reproduce and distribute reprints for Government purposes notwithstanding any copyright notation herein.
				
		\noindent \textbf{Disclosures}: The authors declare no competing financial interests. T.J.K. is a co-founder and shareholder of DEEPLIGHT SA, a startup commercializing PIC-based frequency-agile, low noise lasers.
				
		\noindent \textbf{Data Availability Statement}: The code and data used to produce the plots within this work will be released on the repository \texttt{Zenodo} upon publication of this preprint.
		
		\noindent \textbf{Correspondence and requests for materials} should be addressed to T.J.K. or S.B.
	\end{footnotesize}
\bibliography{citations}
	
\end{document}


\title{Supplementary notes for: Narrow-linewidth, piezoelectrically tunable photonic integrated blue laser}

\author{Anat Siddharth$^{1,2}$,
				Asger B. Gardner$^{3}$, 
			    Xinru Ji$^{1,2}$,
                Shivaprasad U. Hulyal$^{1,2}$,
                Mikael S. Reichler$^{1,2}$, 
			    Alaina Attanasio$^{4}$, 
                Johann Riemensberger$^{1,2}$,
			    Sunil A. Bhave$^{4}$, 
			    Nicolas Volet$^{3}$,
			    Simone Bianconi$^{1,2,\dag}$,
				and Tobias J. Kippenberg$^{1,2,\ddag}$}
\affiliation{
$^1$Laboratory of Photonics and Quantum Measurements, Swiss Federal Institute of Technology Lausanne (EPFL), CH-1015 Lausanne, Switzerland\\
$^2$Center of Quantum Science and Engineering, EPFL, CH-1015 Lausanne, Switzerland\\
$^3$Aarhus University, 8200 Aarhus N, Denmark\\
$^4$Purdue University, West Lafayette, IN-47901, United States\\
}

\setcounter{equation}{0}
\setcounter{figure}{0}
\setcounter{table}{0}

\setcounter{subsection}{0}
\setcounter{section}{0}
\setcounter{secnumdepth}{3}

\maketitle
{\hypersetup{linkcolor=blue}\tableofcontents}
\newpage

\section{State-of-the-Art performance of Photonic Integrated Circuits in the UV–blue spectrum}

\begin{figure}[htb]
	\centering
	\includegraphics[width=\textwidth]{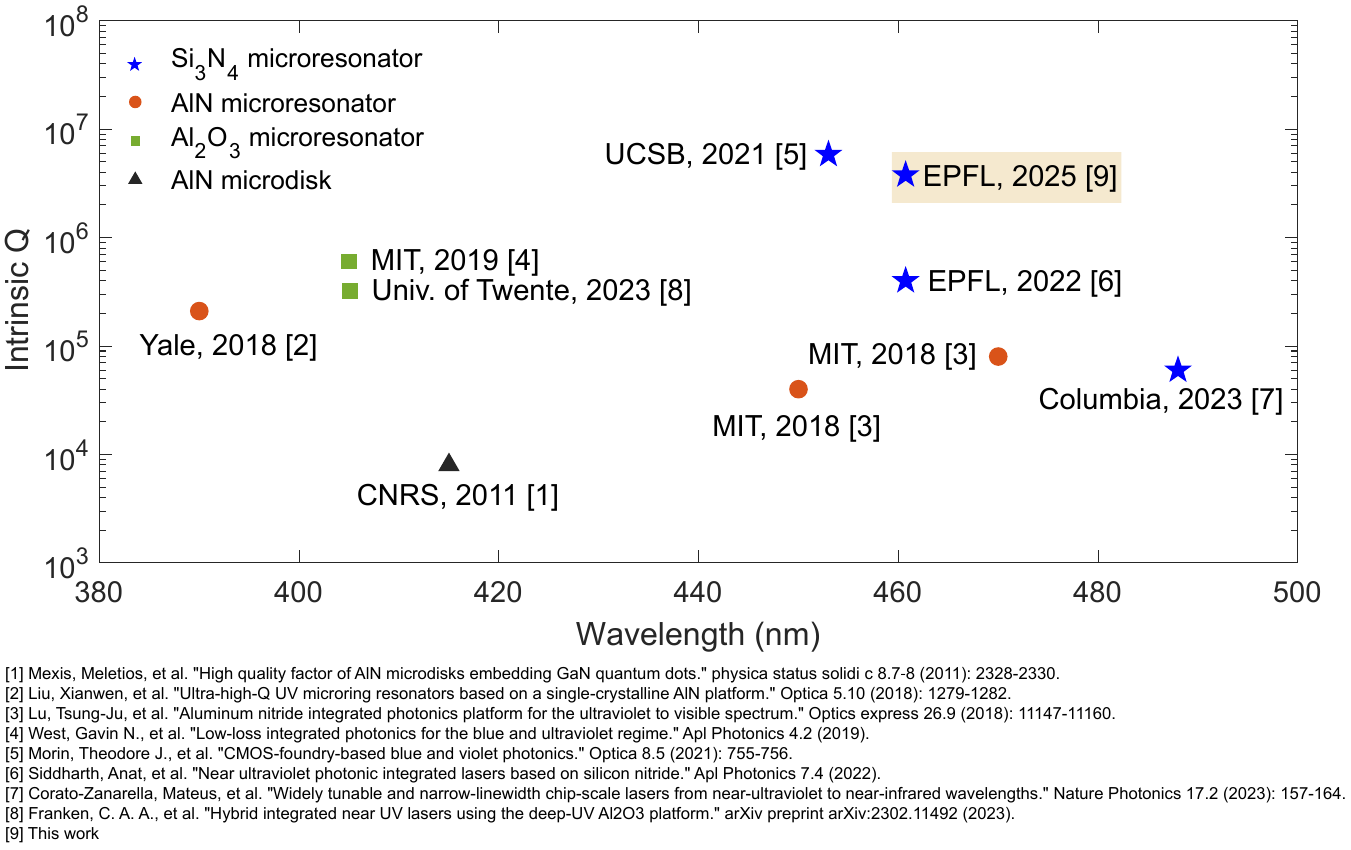}
	\caption{ 
		Comparison of intrinsic quality factor of UV–Blue microresonators across different material platforms.
	}  
	\label{Fig:loss}
\end{figure}
Figure~\ref{Fig:loss} presents a comparison of intrinsic quality factors (Q) as a function of wavelength for various photonic microresonator platforms operating in the ultraviolet-to-blue spectral range (380--500 nm). The reported values include results from several key material systems, namely Si$_3$N$_4$, alumina (Al$_2$O$_3$), and AlN, spanning both microresonator and microdisk geometries. A general trend emerges, showing that higher Q values are typically achieved at longer wavelengths, with values near 450--470 nm reported by UCSB (2021) and the present work (EPFL, 2025). In contrast, microresonators operating near 400 nm demonstrate good Q values, particularly in the Al$_2$O$_3$ and AlN platforms. These results highlight the ongoing effort to push integrated photonics deeper into the blue and UV spectral regions, where high-Q resonators are essential for applications in precision metrology and quantum optics.

\section{Modeling of scattering loss in a waveguide}\label{app:scattering}

Scattering loss in an optical waveguide arises from imperfections such as surface roughness, material inhomogeneities, and sidewall roughness, which cause light to scatter out of the guided mode. 
The primary mechanism of scattering loss in such waveguides is Rayleigh scattering, where the loss scales inversely with the fourth power of wavelength $\lambda^{-4}$, meaning it becomes more severe at shorter wavelengths. 
In microresonators, scattering loss directly affects the quality factor.
Higher scattering loss reduces the Q-factor, thereby increasing the linewidth and lowering the finesse of the resonator. 
This has direct implications for applications such as narrow-linewidth lasers, optical frequency combs, and quantum optics, where high-Q resonators are essential. 
The importance of scattering loss at shorter wavelengths (e.g., visible to near-IR) is particularly significant because fabrication-induced roughness remains on the scale of a few nanometers, which becomes comparable to the wavelength of light, increasing scattering. 
Thus, in applications such as visible photonics, biosensing, and quantum information processing, minimizing sidewall roughness through advanced fabrication techniques (e.g., thermal oxidation smoothing, optimized etching processes) is crucial to achieving low-loss waveguides and high-Q resonators.

\subsection*{Mathematical formulation}

The geometry of the planar waveguide used in the derivation of eq. \ref{eq1} is considered as shown in \ref{Fig:schematic}, where surface roughness is characterized by the root mean square (r.m.s.) deviation from the flatness of an ideal, unperturbed waveguide of a given width \cite{payne1994theoretical}. 
This roughness-induced perturbation affects the guided mode, leading to scattering losses, which are analyzed within the given theoretical framework.
In deriving eq. \ref{eq1}, it is assumed that scattering occurs independently at the top, bottom, and sidewall interfaces of the waveguide. The exponential radiation loss coefficient due to surface roughness scattering in a symmetric single-mode waveguide is given by:
\begin{figure*}[htb]
	\centering
	\includegraphics[width=\textwidth]{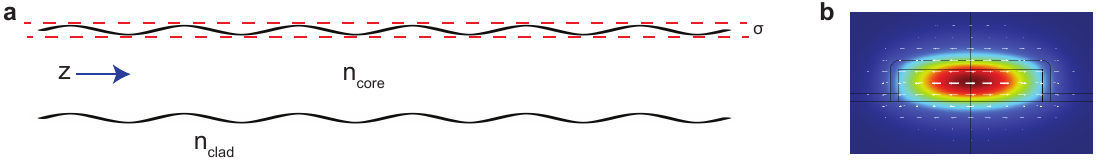}
	\caption{(a) Schematic of the planar waveguide showing the surface roughness described by the r.m.s, deviation $\sigma$ from flatness of an unperturbed waveguide. (b) Optical mode profile of the TE$_{00}$ mode propagating within the planar waveguide.}
	\label{Fig:schematic}
\end{figure*}

\begin{equation}
\alpha = \varphi (\mathrm{n}_{\mathrm{core}}^2 - \mathrm{n}_{\mathrm{clad}}^2)^2 \frac{\mathrm{k}_0^3}{4\pi \mathrm{n}_{\mathrm{core}}} \int_0^\pi \tilde{\mathrm{R}}(\mathrm{k}_0 \mathrm{n}_{\mathrm{eff}} - \mathrm{k}_0 \mathrm{n}_{\mathrm{clad}} \cos\theta) \, d\theta
\label{eq1}
\end{equation}

where
the parameter \( \varphi \) represents the fraction of the total electric field norm confined at the waveguide boundary relative to the entire waveguide cross-section, including the cladding and can be expressed $\varphi = \frac{\int E(x,y) \, dS}{\int\!\!\int E(x,y) \, dA}$.
\(  \mathrm{n}_{\mathrm{eff}} \) and \(  \mathrm{n}_{\mathrm{clad}} \) are the effective refractive index of the optical mode and refractive index of the cladding respectively.
The surface roughness can be statistically characterized using the spectral density function $\tilde{\mathrm{R}}(\Omega)$, which describes how surface roughness variations are distributed across different spatial frequencies.

Mathematically, the spectral density function is related to the autocorrelation function $\mathrm{R(u)}$ of the surface roughness via the Fourier transform
\begin{equation}
\tilde{\mathrm{R}}(\Omega) = \int_{-\infty}^{\infty} \mathrm{R(u) e}^{-i 2\pi \Omega \mathrm{u}} \mathrm{du}
\end{equation}

where
\( \mathrm{R(u)} \) is the autocorrelation function, which quantifies how surface height variations are correlated over a spatial separation u.
A rapid decay of \( \mathrm{R(u)} \) indicates a highly irregular surface, while a slow decay implies smoother variations.
The spectral density function \( \tilde{\mathrm{R}}(\Omega) \) breaks down the roughness into its spatial frequency components. Higher frequencies correspond to fine-grained roughness features, whereas lower frequencies represent large-scale undulations.
Since optical scattering is strongly influenced by spatial frequency components of the roughness that match the guided mode propagation constant, higher spatial frequencies contribute more to scattering loss, especially for shorter wavelengths.

A widely accepted model for characterizing the surface roughness statistics of a waveguide is the exponential autocorrelation function \cite{payne1994theoretical}. Accordingly, we define the autocorrelation function as

\begin{equation}
    \mathrm{R}(u) = \sigma^2 \exp\left(-\frac{|\mathrm{u}|}{\mathrm{L_c}}\right)
\end{equation}
where $\sigma^2$ represents the variance of the surface height fluctuations (i.e., the mean square roughness), and $\mathrm{L_c}$  denotes the correlation length, which quantifies the characteristic distance over which surface roughness remains correlated.
If $\mathrm{L_c}$ is large, the surface roughness varies smoothly over a longer distance, meaning the roughness features are more extended.
Short $\mathrm{L_c}$ (high-frequency roughness) contributes more to scattering, especially at shorter wavelengths.

Substituting the exponential autocorrelation function \( \mathrm{R_u} \) into equation \ref{eq1}, the scattering loss at the interface (eg. top surface) of the waveguide is expressed as:

\begin{equation}
    \alpha_{\mathrm{top}} = \frac{\varphi}{\lambda^3 \mathrm{n_{core}}} ( \mathrm{n_{core}}^2 -  \mathrm{n_{clad}}^2) \mathrm{S_{top}} \cdot 2\pi^2\cdot10\cdot\mathrm{log_{10}e} [\text{dB/m}].
\end{equation}

where \( \mathrm{S_{top}} \) is defined as:

\begin{equation}
    \mathrm{S_{top}} = \sqrt{2} \sigma_{\mathrm{top}}^2 \mathrm{L}_{\mathrm{c_{top}}} \pi \frac{\sqrt{\sqrt{4 \mathrm{k}_0^2 \mathrm{n}_{\mathrm{eff}}^2 \mathrm{L}_{\mathrm{c_{top}}}^2 + \mathrm{K}^2} + \mathrm{K}}}{\sqrt{4 \mathrm{k}_0^2 \mathrm{n}_{\mathrm{eff}}^2 \mathrm{L}_{\mathrm{c_{top}}}^2 + \mathrm{K}^2}}.
\end{equation}

where \( \mathrm{K} \) is given by:

\begin{equation}
    \mathrm{K} = 1 - \mathrm{L}_{\mathrm{c_{top}}}^2 \mathrm{k}_0^2 (\mathrm{n}_{\mathrm{eff}}^2 -  \mathrm{n_{clad}}^2).
\end{equation}

Similarly, the scattering loss at the sidewall interfaces can also be calculated.

This formulation models the scattering-induced radiation loss by integrating over all possible scattering angles, assuming that the roughness-induced perturbations are sufficiently small to be treated within a perturbative framework. 
The equation highlights the dependence of scattering loss on the waveguide's refractive index contrast, the spatial frequency content of the roughness, and the mode distribution at the interface.

\begin{figure*}[htb]
	\centering
	\includegraphics[width=\textwidth]{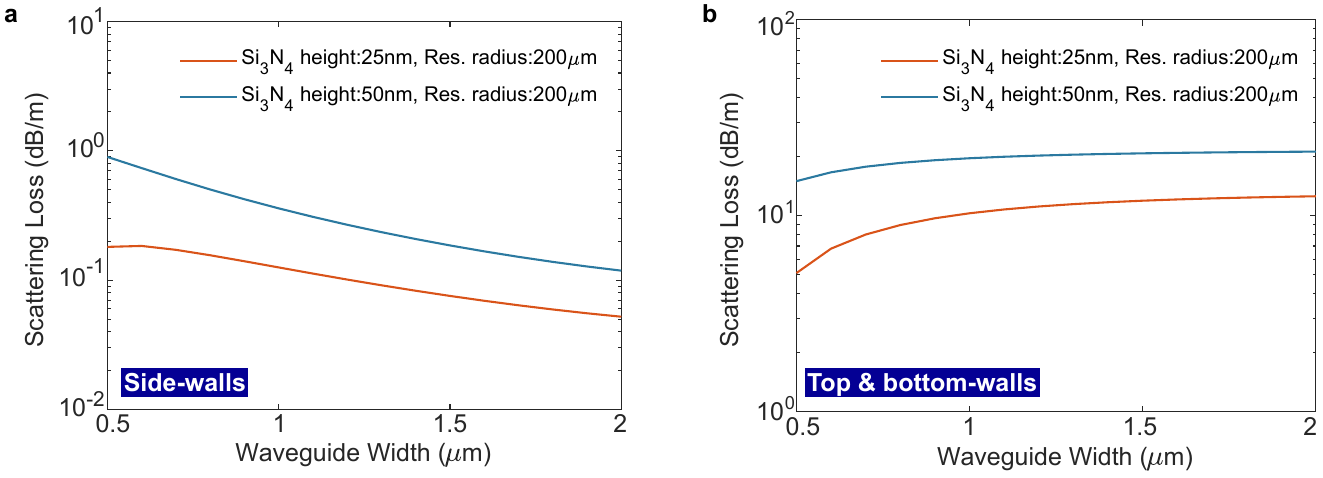}
	\caption{(a) Sidewall scattering loss and (b) top/bottom surface scattering loss as a function of waveguide width for 25 nm and 50 nm thin \SiN~ waveguides at 650.31 THz optical frequency.}
	\label{Fig:thinblue}
\end{figure*}
We integrate the aforementioned formulation into a COMSOL-linked MATLAB model to compute the scattering loss for 25 nm and 50 nm thin \SiN waveguides, considering a fixed resonator radius of 200 $\mu$m. 

The results shown in Figure \ref{Fig:thinblue} indicate that scattering losses from both sidewalls and top/bottom interfaces are lower for the 25 nm thin \SiN waveguide compared to the 50 nm counterpart at 650.31 THz optical frequency. 
Additionally, the analysis reveals that sidewall scattering decreases as the waveguide width increases for a fixed height. This reduction occurs due to stronger optical mode confinement within the waveguide core, leading to reduced interaction with the sidewalls. 
Conversely, top/bottom surface scattering exhibits an increasing trend with waveguide width, as a larger waveguide width exposes more of the optical mode to the rough top and bottom surfaces, thereby enhancing scattering losses.
These findings highlight the trade-offs between waveguide geometry and scattering-induced propagation loss, which are crucial for optimizing low-loss visible photonic integrated circuits.

\section{Numerical analysis of self-injection locking}

\begin{figure}[htb]
    \centering
    \includegraphics[width=0.5\linewidth]{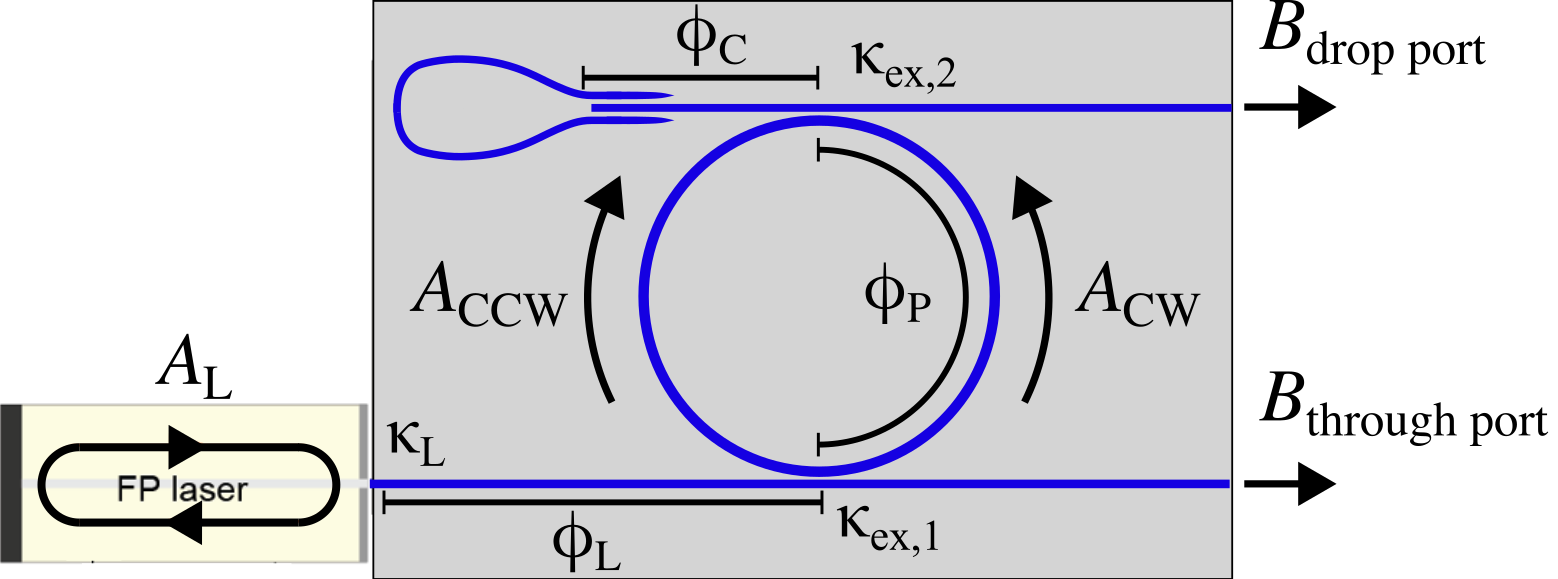}
    \caption{Schematic of the laser-resonator system with a mirror-assisted drop port.}
    \label{fig:sil_system}
\end{figure}

The system consisting of a laser, described by the intracavity field amplitude $A_\mathrm{L}$ and carrier density $N$, coupled to a microresonator with clockwise and counterclockwise fields $A_\mathrm{CW}$ and $A_\mathrm{CCW}$, can be described by the following system of equations \cite{kondratiev2017self,lihachev2022low,galiev2021mirror}:
\begin{align}\label{eqs1}
    \frac{d N}{d t} &= \frac{I_\mathrm{bias} + \Delta I}{e V} - \kappa_N N - g V (N-N_0)|A_L|^2,\\
    \frac{d A_L}{d t} &= \bigl[ (1 + i\alpha)(N-N_0)g V - \frac{\tilde{\kappa}_L}{2} + i\delta\omega_L + i \zeta \Delta I \bigr] A_L + \sqrt{\kappa_{\mathrm{ex},1}\kappa_{\mathrm{ex},L}\frac{\tau_R}{\tau_L}}e^{i\phi_L}A_\mathrm{CW} + F_A,\\
    \frac{d A_\mathrm{CW}}{d t} &= \bigl[i\delta\omega_R - \frac{\tilde{\kappa}_R}{2} - i\Gamma(|A_\mathrm{CW}| + 2|A_\mathrm{CCW}|)\bigr]A_\mathrm{CW} + i\frac{\kappa_\mathrm{sc}}{2}A_\mathrm{CCW} + \sqrt{\kappa_{\mathrm{ex},1}\kappa_{\mathrm{ex},L}\frac{\tau_L}{\tau_R}} e^{i\phi_L}A_L,\\
    \frac{d A_\mathrm{CCW}}{d t} &= \bigl[i\delta\omega_R - \frac{\tilde{\kappa}_R}{2} - i\Gamma(|A_\mathrm{CCW}| + 2|A_\mathrm{CW}|)\bigr]A_\mathrm{CCW} + \bigl(i\frac{\kappa_\mathrm{sc}}{2} - R_M\kappa_{\mathrm{ex},2}e^{i(2\phi_P + 2\phi_C)}\bigr)A_\mathrm{CW}\label{eqs9}
\end{align}

In the laser rate equations, $I_\mathrm{bias}$ is the biased injection current, $\Delta I$ the injection current modulation, $e$ is the electron charge, $V $ is the volume of the active region, $\kappa_N$ is the carrier recombination rate, $g$ is the carrier gain coefficient, $N_0$ is the carrier density at transparency, $\alpha$ is the linewidth enhancement factor, $\tilde{\kappa} = \kappa_\mathrm{L} + \kappa_\mathrm{ex,L}$ includes the intrinsic laser cavity loss rate $\kappa_\mathrm{L}$ and the external coupling rate $\kappa_\mathrm{ex,L}$, $\delta\omega_\mathrm{L}$ is the laser detuning, $\zeta$ is the current-frequency tuning coefficient, $\tau_\mathrm{L}$ is the round-trip time of the laser cavity, and $F_A$ is a Langevin noise term describing spontaneous emission into the lasing field. The noise in carrier density fluctuations can be ignored as it's contribution is negligible compared to the noise in the optical field~\cite{ohtsuboChaos}. In the coupled resonator mode equations, $\delta \omega_\mathrm{R}$ is the resonator detuning, $\tilde{\kappa} = \kappa_\mathrm{0} + \kappa_\mathrm{ex,1} + \kappa_\mathrm{ex,2}$ is the loaded cavity loss rate including the intrinsic loss rate $\kappa_\mathrm{0}$, external coupling to the bus waveguide $\kappa_\mathrm{ex,1}$ and external coupling to the drop port $\kappa_\mathrm{ex,2}$, $\Gamma$ is the Kerr frequency shift coefficient, $\kappa_\mathrm{sc}$ is the coupling between the CW and CCW modes, $\tau_\mathrm{R}$ is the round trip time of the cavity, $B_\mathrm{in}$ is the forward field amplitude in the bus waveguide, $R_\mathrm{m}$ is the reflectivity of the drop port mirror, and $\phi_\mathrm{P}$ and $\phi_\mathrm{C}$ are the phase delays between the bus and drop coupler in the resonator and the drop port and the mirror, respectively. The Kerr frequency shift coefficient is given by $\Gamma = \frac{\hbar \omega_0^2 c n_2}{n_0^2}$, where $\hbar$ is the reduced Planck constant, $\omega_0$ is the optical frequency, $c$ is the speed of light, $n_2$ is the Kerr nonlinear index, and $n_0$ is the refractive index. The experimentally observable field amplitudes at the through and drop ports are given by
\begin{align}
    B_{\textrm{through port}} &= i \sqrt{\kappa_{\mathrm{ex},L}\tau_L(1-\kappa_{\mathrm{ex},1}\tau_R)}e^{i\phi_L}A_L + i\sqrt{\kappa_{\mathrm{ex},1}\tau_R}A_\mathrm{CW},\\
    B_{\textrm{drop port}} &= iR_m\sqrt{\kappa_{\mathrm{ex},2}\tau_R(1-\kappa_{\mathrm{ex},2}\tau_R)}e^{i(\phi_P+2\phi_C)}A_\mathrm{CW} + i\sqrt{\kappa_{\mathrm{ex},2}\tau_R}e^{-i\phi_P}A_\mathrm{CCW}.\\
\end{align}

The Langevin noise term $F_\mathrm{A}$ describing spontaneous emission can be considered to originate from a white noise process with a Gaussian probability distribution \cite{ohtsuboChaos}. Numerically, $F_\mathrm{A}$ can thus be implemented in the rate equations as
\begin{align}
    F_\mathrm{A} = \xi \sqrt{\frac{2 R_\mathrm{sp} N}{\Delta t}},
\end{align}
where $\xi$ is a zero-mean, unit-variance Gaussian distribution, $R_\mathrm{sp}$ is the spontaneous emission coefficient, $N$ is the carrier density, and $\Delta t$ is the time step used in the numerical solver. The Langevin noise term is used to initiate stimulated optical amplification in the absence of an initial field.
\begin{table}[h]
\caption{Parameters used in the numerical simulations.}
\label{tab:pars}
\begin{center}
\begin{tabular}{ |c|c|c|c| } 
 \hline
 Symbol & Value & Unit & Definition\\
 \hline
 $\hbar$ & $1.05 \times 10^{-34}$& $\mathrm{m}^{2}\mathrm{kg}$ $\mathrm{s}^{-1}$ & Reduced Planck constant \\
 $e$ & $1.6 \times 10^{-19}$& C & Elementary charge \\
 $c$ & $3 \times 10^{8}$ & $\mathrm{m}$ $\mathrm{s}^{-1}$ & Speed of light \\
 $\omega_0$ & $6.5 \times 10^{14}$ & $\mathrm{s}^{-1}$ & Optical frequency \\
 $\kappa_\mathrm{N}$ & $1 \times 10^{9}$& $\mathrm{s}^{-1}$ & Carrier recombination rate \\ 
 $N_0$ & $1 \times 10^{24}$& $\mathrm{m}^{-3}$ & Carrier density at transparency \\ 
 $I_\mathrm{bias}$ & $0.25$& A & Injection current bias \\ 
 $g$ & $1 \times 10^{6}$&$\mathrm{m}^{3}\mathrm{s}^{-1}$& Carrier gain coefficient \\
 $\alpha$ & 5& & Linewidth enhancement factor \\
 $\kappa_\mathrm{L}$ & $2 \times \pi \times 5 \times 10^{10}$& rad $\mathrm{s}^{-1}$ & Laser intrinsic loss rate \\
 $\kappa_\mathrm{ex,L}$ & $2 \times \pi \times 5 \times 10^{9}$& rad $\mathrm{s}^{-1}$ & Laser external coupling rate \\
 $\zeta$ &$5 \times 10^{11}$ & $\mathrm{s}^{-1} \mathrm{A}^{-1}$ & Current-frequency tuning coefficient \\ 
 $V$ &$2 \times 10^{-16}$ & $\mathrm{m}^{3}$ & Volume of the active section \\ 
 $R_\mathrm{sp}$ & $1 \times 10^{-10}$& $\mathrm{s}^{-1}$ & Spontaneous emission coefficient \\
 $\tau_\mathrm{L}$ & $1 \times 10^{-12}$& s & Laser cavity roundtrip time \\
 $\kappa_\mathrm{0}$ & $2 \times \pi \times 1690 \times 10^6$& rad $\mathrm{s}^{-1}$ & Resonator intrinsic loss rate \\ 
 $\kappa_\mathrm{ex,1}$ & $2 \times \pi \times 320 \times 10^6$& rad $\mathrm{s}^{-1}$ & Resonator-bus coupling rate \\
 $\kappa_\mathrm{ex,2}$ & $2 \times \pi \times 320 \times 10^6$& rad $\mathrm{s}^{-1}$ & Resonator-drop-port coupling rate \\
 $\kappa_\mathrm{sc}$ & $2 \times \pi \times 100 \times 10^6$& rad $\mathrm{s}^{-1}$ & Coupling rate between CW and CCW resonator modes \\
 $\tau_\mathrm{0}$ & $1 \times 10^{-11}$& s & Resonator roundtrip time \\
 $R_\mathrm{m}$ & $1$&  & Drop port mirror reflectivity \\
 $n_0$ & 2&  & Microresonator refractive index \\
 $n_2$ & $3 \times 10^{-20}$& $\mathrm{m}^2 \mathrm{W}$ & Effective nonlinear refractive index \\
 $\phi_\mathrm{P}$ & $0$& rad & Through port - drop port phase delay \\
 $\phi_\mathrm{C}$ & $-\pi/4$& rad & Drop port - mirror phase delay \\
 $\phi_\mathrm{L}$ & 1.5& rad & Laser - resonator phase delay \\
  \hline
\end{tabular}
\end{center}
\end{table}

\begin{figure}[htb]
\centering
\begin{subfigure}{.25\textwidth}
  \centering
  \includegraphics[width=1\linewidth]{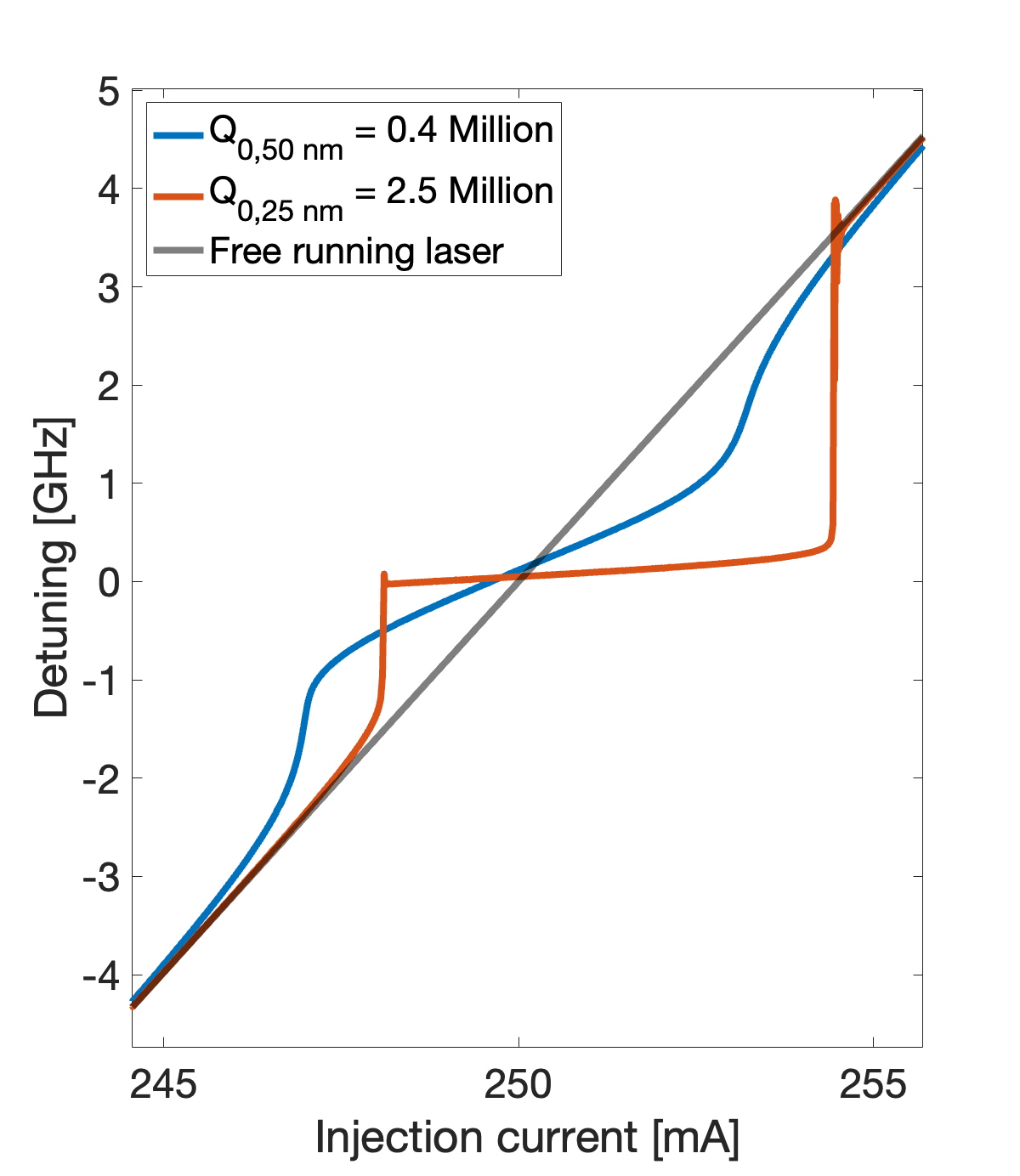}
\end{subfigure}%
\begin{subfigure}{.25\textwidth}
  \centering
  \includegraphics[width=1\linewidth]{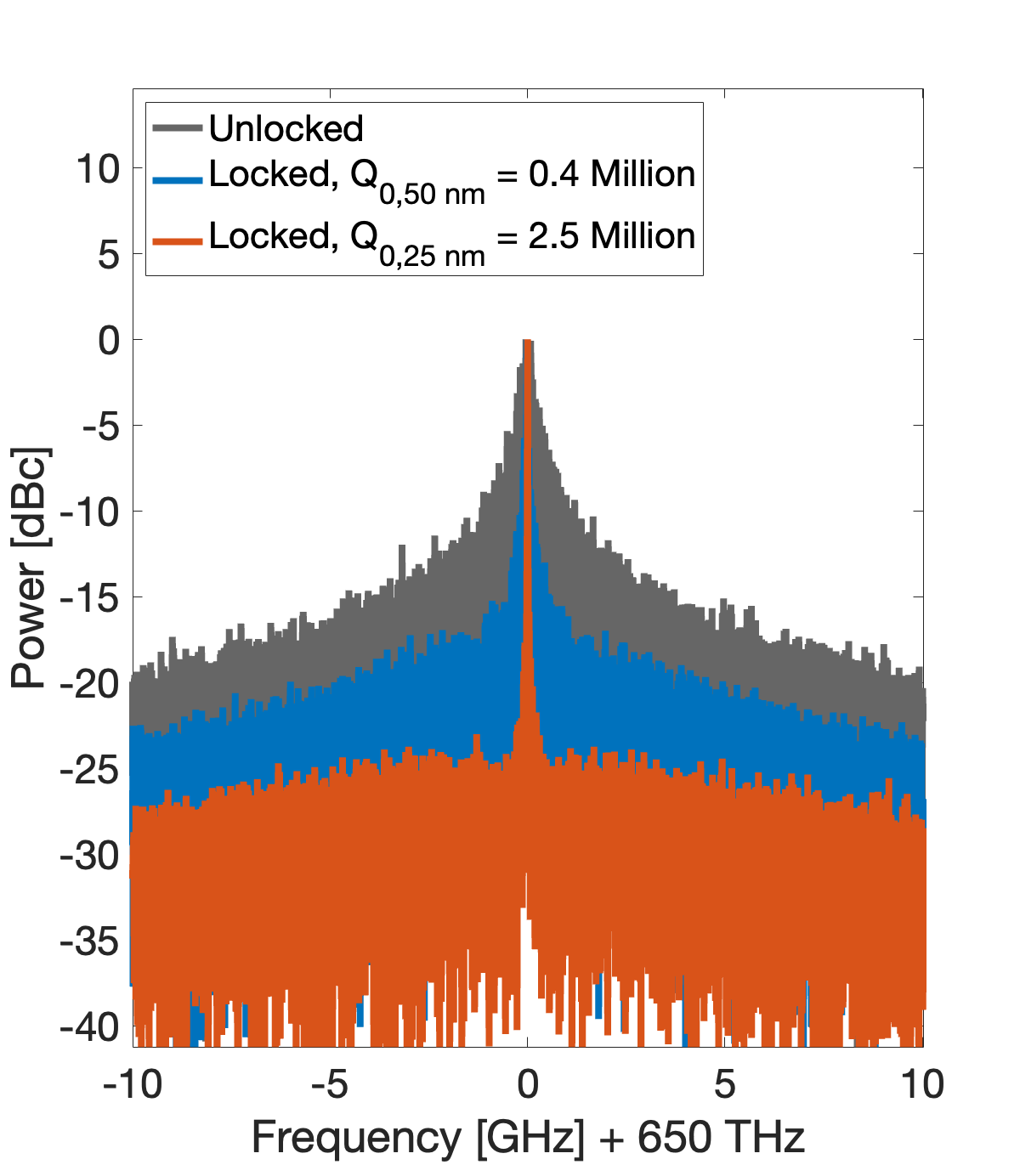}
\end{subfigure}
\begin{subfigure}{.25\textwidth}
  \centering
  \includegraphics[width=1\linewidth]{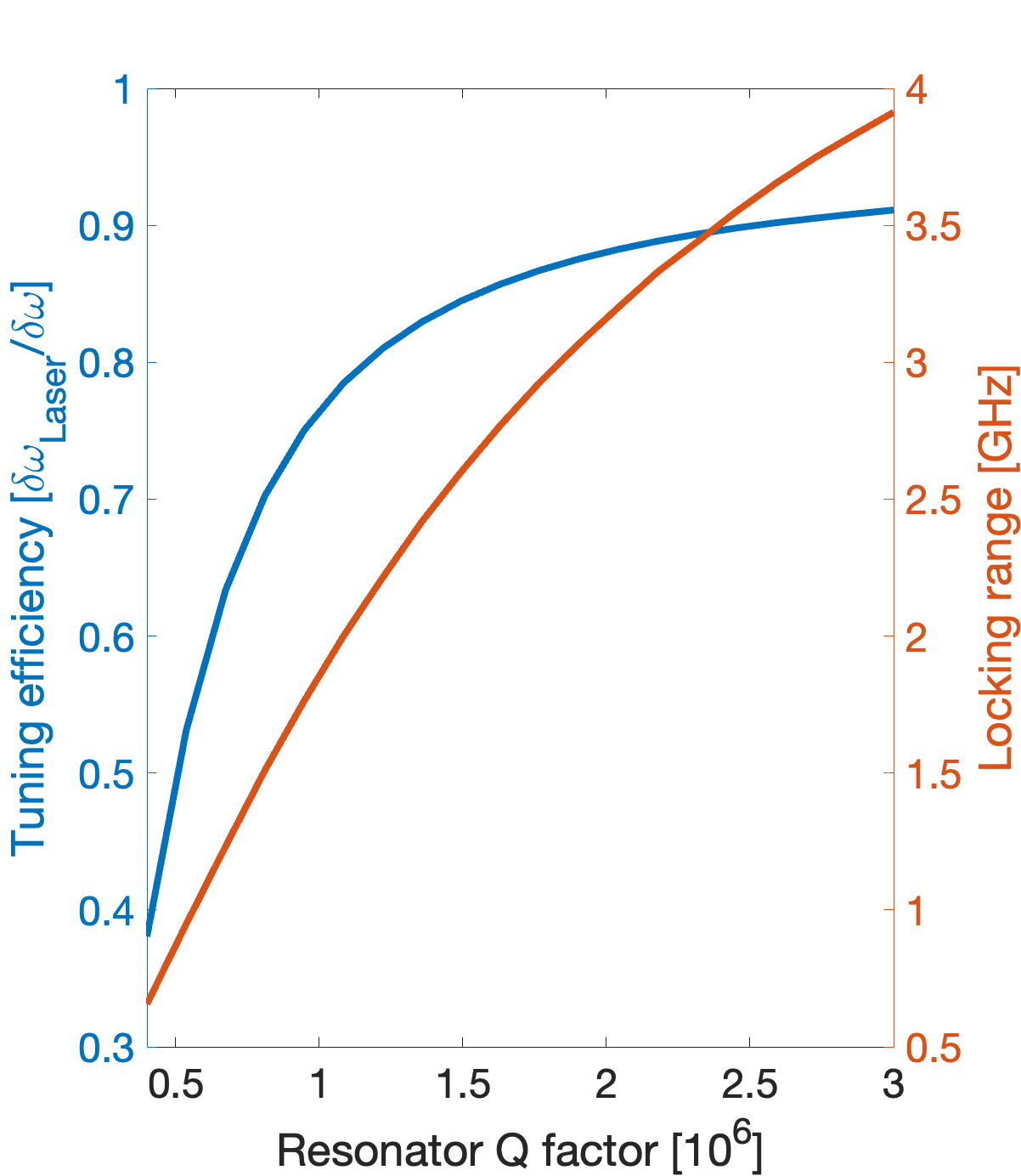}
\end{subfigure}%
\caption{\textbf{Effect of the intrinsic resonator Q-factor on self-injection locking.} Left: Laser frequency detuning from the microcavity resonance as the laser is tuned via current modulation. Middle: Optical power spectra simulated using an exaggerated $R_\mathrm{sp}$ of $10^6$ s$^{-1}$. Right: Tuning efficiency and locking range as a function of intrinsic resonator Q-factor, where the laser outcoupling is set to 1 GHz.}
\label{fig:SI_SIL_1}
\end{figure}

We demonstrate the effect of the intrinsic quality factor of the microresonator on self-injection locking by numerically integrating Eqs.~\ref{eqs1}-\ref{eqs9} with a Runge-Kutta method and using parameters shown in Table~\ref{tab:pars}. Fig.~\ref{fig:SI_SIL_1} clearly shows the increase in stabilization coefficient and reduction in linewidth obtainable with the higher quality factor. The tuning efficiency and locking range shown in Fig.~\ref{fig:SI_SIL_1} show a significant increase as the quality factor is increased. 
\begin{figure}[htb]
\centering
\begin{subfigure}{.33\textwidth}
  \centering
  \includegraphics[width=1\linewidth]{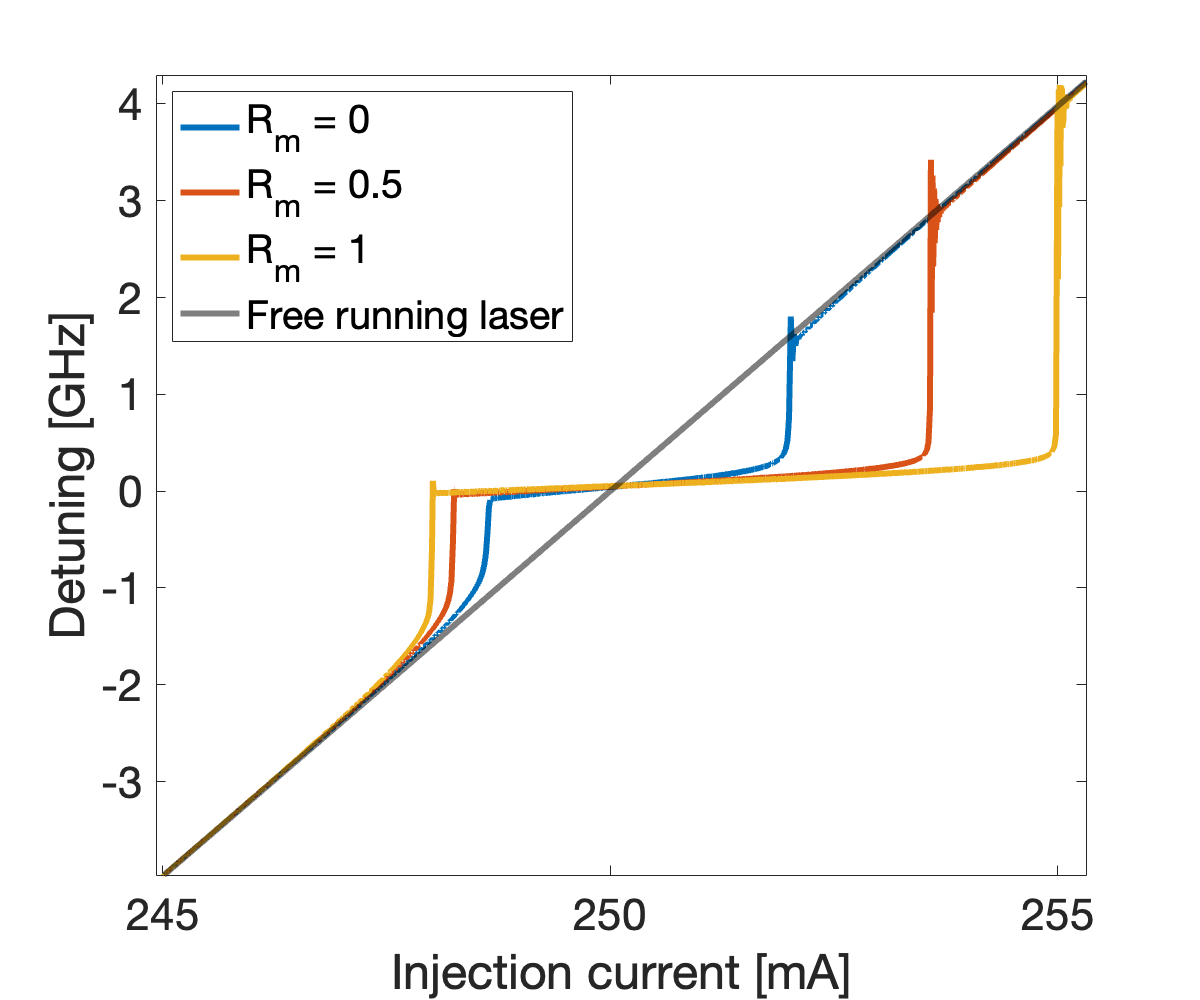}
\end{subfigure}%
\begin{subfigure}{.33\textwidth}
  \centering
  \includegraphics[width=1\linewidth]{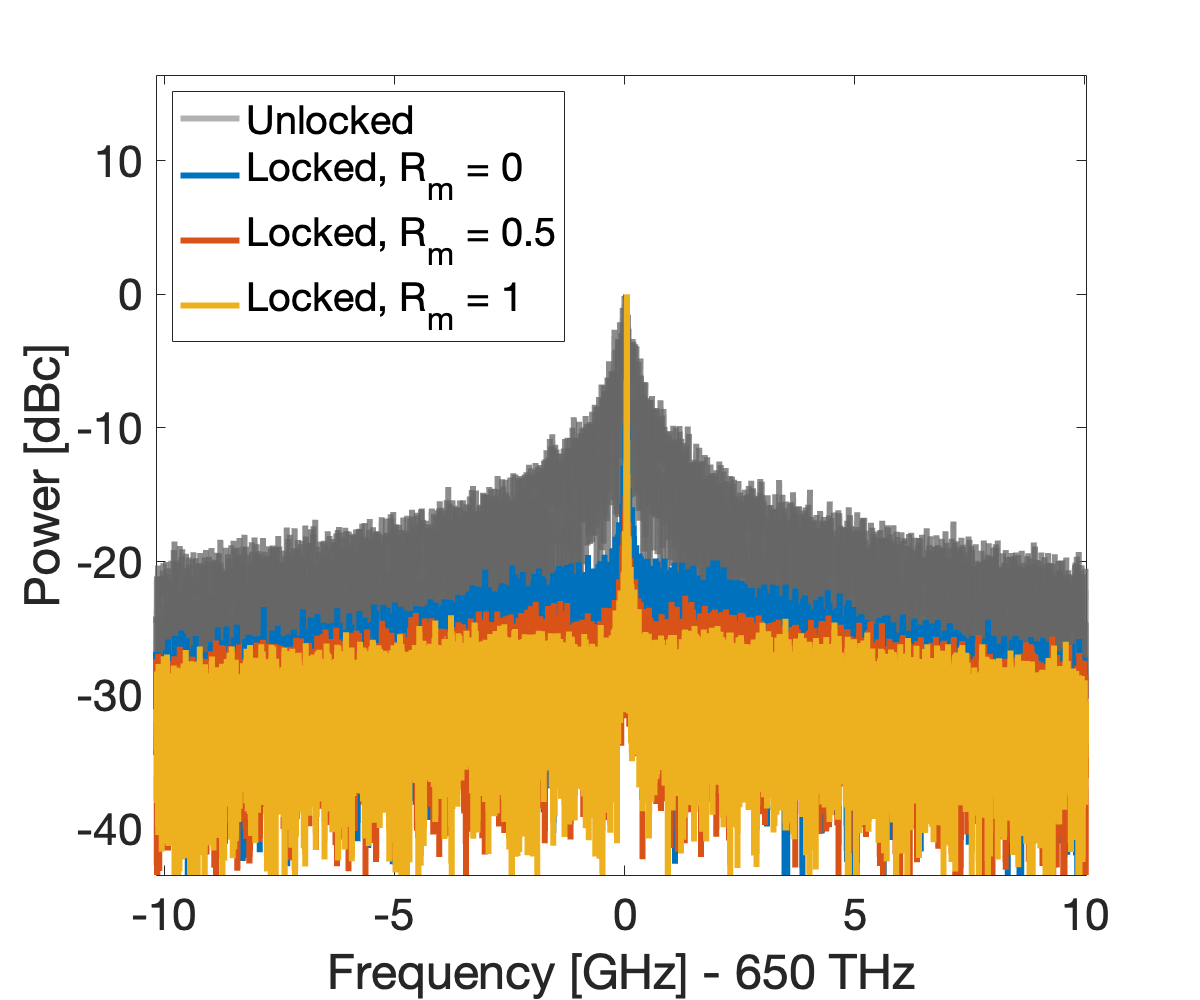}
\end{subfigure}
\caption{\textbf{Effect of the drop port mirror reflectance on self-injection locking.} Left: Laser frequency detuning from the microcavity resonance as the laser is tuned via current modulation. Right: Optical power spectra simulated using an exaggerated $R_\mathrm{sp}$ of $10^6$ s$^{-1}$.}
\label{fig:SI_SIL_3}
\end{figure}

To investigate the effect of increased drop port mirror losses in the 25 nm $\mathrm{Si}_3\mathrm{N}_4$ platform, we simulate laser generation frequency curves and optical power spectra for different drop port mirror reflectivities in Fig.~\ref{fig:SI_SIL_3}. Here, the intrinsic Q-value is set to 2.5 million, and the Rayleigh backscattering rate is set equal to the intrinsic loss rate. The reduced optical feedback shortens the locking range by up to 70 \% compared to the case with unity drop port mirror reflection, as only the residual Rayleigh backscattering provides optical feedback from the resonator to the laser. The weaker optical feedback also reduces the stabilization coefficient, leading to a increased Lorentzian linewidth.
\begin{figure}[htb]
\centering
\begin{subfigure}{.5\textwidth}
  \centering
  \includegraphics[width=1\linewidth]{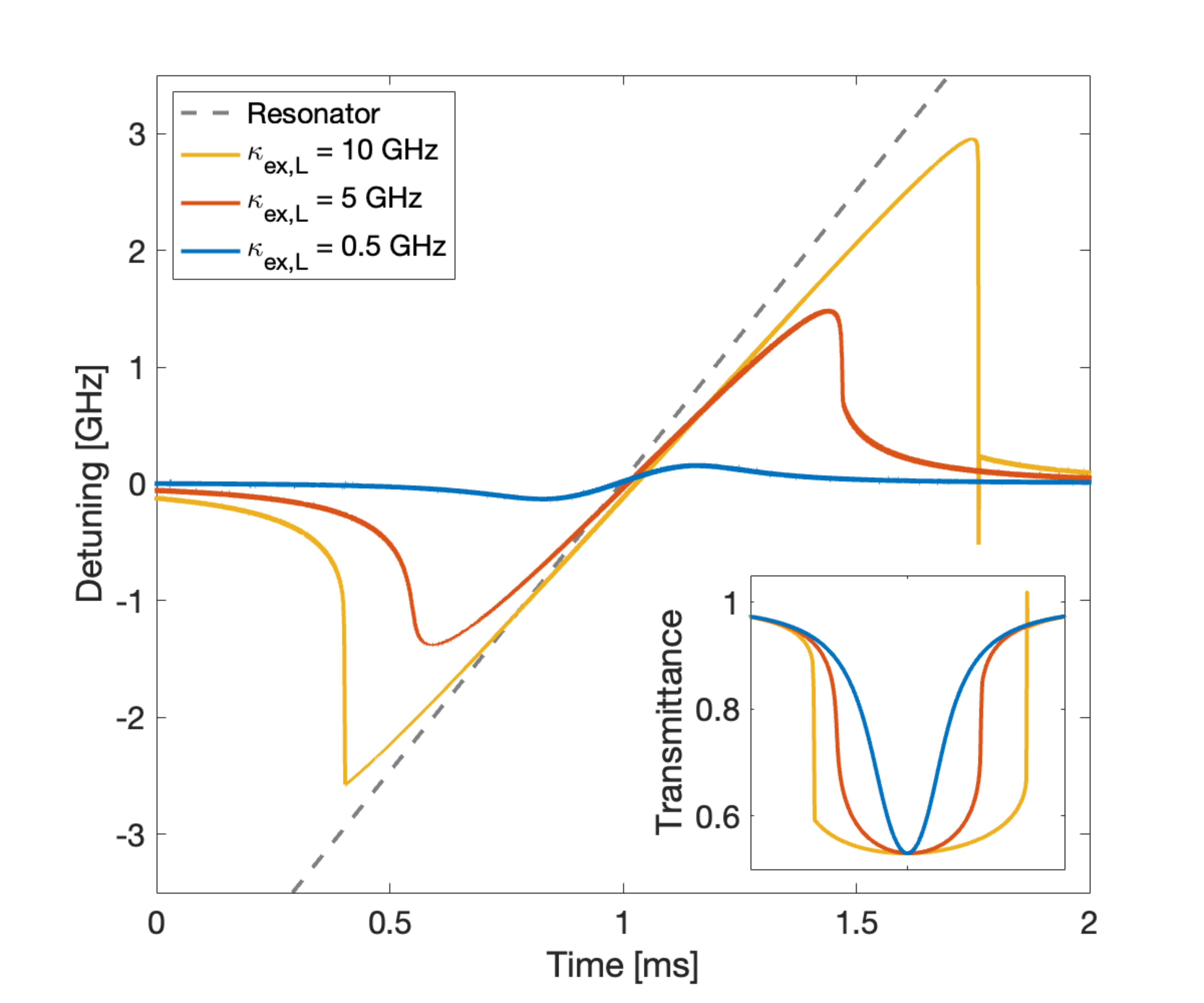}
\end{subfigure}%
\begin{subfigure}{.5\textwidth}
  \centering
  \includegraphics[width=1\linewidth]{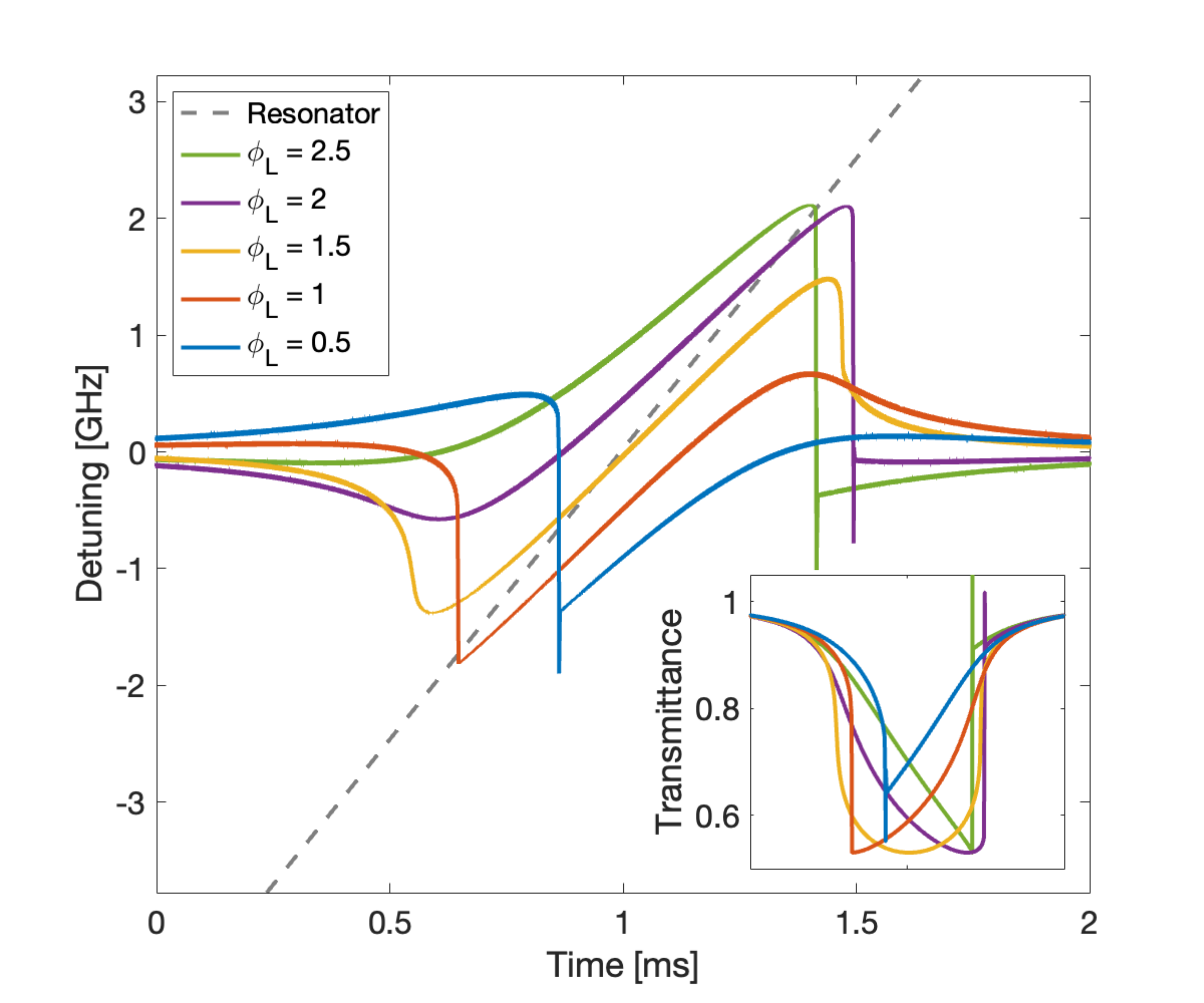}
\end{subfigure}
\caption{\textbf{Effect of the laser external coupling rate and feedback phase on self-injection locking.} Laser frequency evolution at different values of $\kappa_\mathrm{ex,L}/2 \pi$ and $\phi_\mathrm{L}$ as the microresonator resonance is tuned through the free running laser frequency. On the left, the feedback phase $\phi_\mathrm{L}$ is kept at a constant 1.5, and on the right, the laser outcoupling $\kappa_\mathrm{ex,L}/2\pi$ is kept at 5 GHz.}
\label{fig:SI_SIL_2}
\end{figure}

In Fig.~\ref{fig:SI_SIL_2}, we simulate laser generation frequency curves for various values of $\kappa_\mathrm{ex,L}$ and $\phi_\mathrm{L}$ as the resonance of the resonator is tuned through the free running laser frequency. Whereas the laser outcoupling greatly influences the stabilization coefficient due to variations in backreflection strenght, feedback phase deviations have a more subtle effect on the stabilization coefficient, tuning efficiency, and linewidth reduction, as small perturbations from the optimal value mainly shift the laser generation frequency within the linewidth of the microresonator

\section{Comparison of bending losses in 25 nm and 50 nm thick \SiN~platforms at 461 nm}

\begin{figure}[htb]
	\centering
	\includegraphics[width=0.8\textwidth]{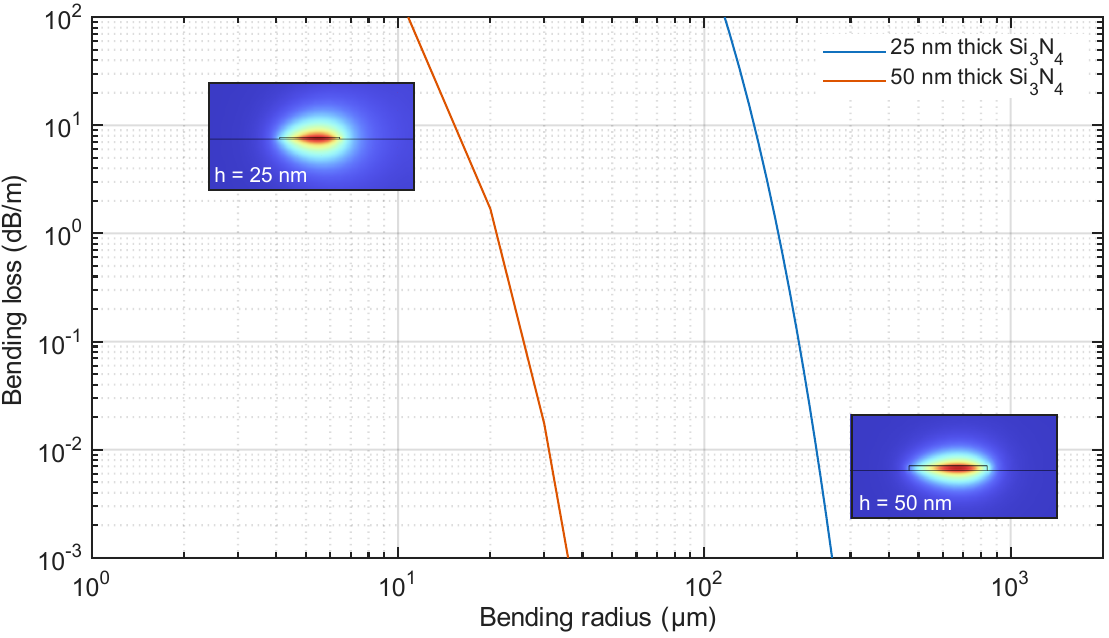}
	\caption{ 
		Comparison of optical bending loss at different bending radii of the \SiN waveguide in the 25 nm and the 50 nm thick \SiN platform.
	}  
	\label{Fig:BendingLoss}
\end{figure}
Figure~\ref{Fig:BendingLoss} shows the simulated bending loss (in dB/m) as a function of bend radius for Si$_3$N$_4$ waveguides with thicknesses of 25 nm (blue curve) and 50 nm (red curve). The loss increases sharply as the bend radius decreases, with the 25 nm platform exhibiting significantly higher bending loss across the entire range.

This disparity arises from differences in optical mode confinement, the 25 nm waveguide supports a confinement factor of only $\sim$9\%, whereas the 50 nm platform provides $\sim$30\% confinement. Weaker confinement in the thinner film causes greater mode leakage when bent, thus enhancing radiation loss. The insets show the corresponding fundamental mode profiles, clearly illustrating the difference in confinement and field extension into the cladding.

\section{Frequency-modulated continuous wave operation of SIL blue laser based on 25 nm thick \SiN~platform}

\begin{figure}[htb]
	\centering
	\includegraphics[width=0.8\textwidth]{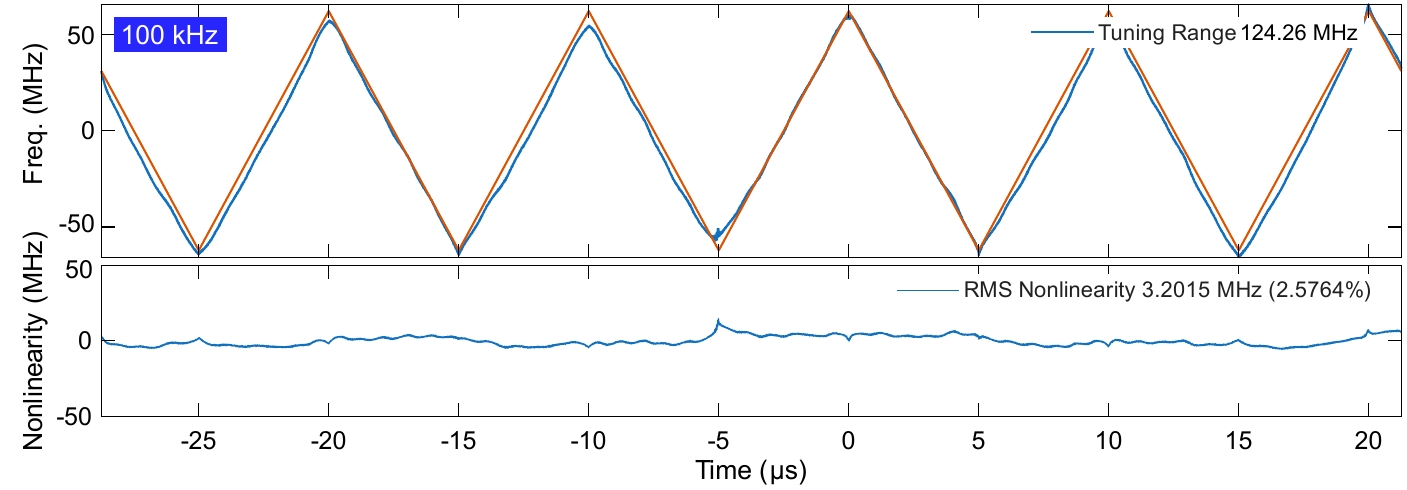}
	\caption{ 
		The SIL laser (25 nm thick \SiN) is modulated with a 100 kHz at 10 V$_{\text{pp}}$ applied accross the AlN piezoactuator. The nonlinearity is determined by fitting an ideal triangular waveform to the tuning curve obtained via Hilbert transform of the iMZI output.
	}  
	\label{fig7}
\end{figure}

The reduced tuning range in the 25~nm \SiN~device, as compared to the 50~nm counterparts, is attributed to a narrow injection locking range, which constrains the laser's dynamic frequency pulling ability. 
A narrower locking range indicates weaker optical feedback from the microring resonator to the gain section.
A key contributing factor is likely non-optimized back-reflection from the loop mirror, potentially due to tight waveguide bends in the loop.
For ultra-thin \SiN~(25~nm), bending radii are more constrained due to lower confinement, which leads to higher bending loss and reduced constructive interference at the reflection interface.
Consequently, insufficient optical feedback weakens the phase stabilization required for robust locking and broad tuning.
Importantly, this limitation is not fundamental to the 25~nm \SiN~platform.
Improved loop mirror designs, with relaxed bend radii could significantly enhance the back-reflection and broaden the locking range.


\bibliographystyle{apsrev4-2}
\bibliography{refs}